\documentclass[twocolumn,pra]{revtex4}
\usepackage{graphicx}
\usepackage{color}
\usepackage{amsmath}
\usepackage{hyperref}
\usepackage{verbatim}
\usepackage{graphicx}
\usepackage{color}
\usepackage{xcolor}
\usepackage{ragged2e}
\usepackage{wrapfig}
\usepackage{setspace}
\usepackage{braket}
\usepackage{bbold}

 \definecolor{boxback}{HTML}{FFF8B5}
 
 \definecolor{applegreen}{rgb}{0, 0.5, 0.0}

\definecolor{REDRED}{HTML}{C5232F}

\definecolor{mygreen}{rgb}{0,0.5,0}
\definecolor{myblue}{rgb}{0,0,0.75}
\definecolor{mymagenta}{cmyk}{0,1,0,0.12}

\newcommand{\Hres}{\hat{H}_R}
\newcommand{\Htarg}{\hat{H}_T}
\newcommand{\CPsym}{\hat{CP}}

\newcommand{\Id}{\mathbb{1}}

\newcommand{\paramvec}{\boldsymbol{\theta}}

\begin{document}
\bibliographystyle{apsrev}

\title{Self-Verifying Variational Quantum Simulation of the Lattice Schwinger Model}

\author{C.~Kokail${}^{\ast}$}
\author{C.~Maier${}^{\ast}$}
\author{R.~van Bijnen${}^{\ast}$}
\author{T.~Brydges}
\author{M.~K.~Joshi}
\author{P.~Jurcevic}
\author{C.~A.~Muschik}
\author{P.~Silvi}
\author{R.~Blatt}
\author{C.~F.~Roos}
\author{P.~Zoller}

\affiliation{Center for Quantum Physics, and Institute for Experimental Physics, University of Innsbruck} 
\affiliation{Institute for Quantum Optics and Quantum Information, Austrian Academy of Sciences,
Innsbruck, Austria}

\date{\today}

\begin{abstract}
Hybrid classical-quantum algorithms aim at variationally solving optimisation problems, using a feedback loop between a classical computer and a quantum co-processor, while 
benefitting from quantum resources. Here we present experiments demonstrating self-verifying, hybrid,
variational quantum simulation of lattice models in condensed matter and high-energy physics.  Contrary to analog quantum simulation, this approach forgoes the requirement of realising the targeted Hamiltonian directly in the laboratory, thus allowing the study of a wide variety of previously intractable target models.
Here, we focus on the Lattice Schwinger model, a gauge theory of 1D quantum electrodynamics. Our quantum co-processor is a programmable, trapped-ion analog quantum simulator with up to 20 qubits, capable of generating families of entangled trial states respecting symmetries of the target Hamiltonian. We determine ground states, energy gaps and, by measuring variances of the Schwinger Hamiltonian, we provide algorithmic error bars for energies, thus addressing the long-standing challenge of verifying quantum simulation.
\end{abstract}

\maketitle



The development of quantum simulators provides us with new tools to solve the quantum many-body problem of condensed matter and high-energy physics
\cite{CZ2012,Georgescu2014}.  In analog quantum simulation (AQS), highly controllable quantum systems directly realise specific model Hamiltonians using the natural quantum resources available for a given physical platform \cite{GrossQSim2017,Bernien51atoms,BrowaeysQSim2016,Bollinger2DQsim2012,Monroe53qubits,BlattQSim2012,Houck2012,OberthalerQSim2018}. AQS  can  be  scaled  to  a  large  number  of  particles while  still  maintaining  excellent  quantum  coherence. However, for complex target models, such as lattice gauge theories,
realisation in the framework of AQS becomes increasingly challenging. In contrast, digital quantum simulation (DQS) is able to propagate the many-body wavefunctions for a generic quantum many-body Hamiltonian as Trotterised time evolution on a universal quantum computer \cite{Lloyd96,LanyonDQS2011,Martinez2016,WallraffPRX2017}. Yet, realisation of a large scale, fault-tolerant quantum computer remains a long-term goal, and thus far DQS has remained limited to very small system sizes. 
In the present work, we focus on variational quantum simulation (VQS) as a third way of simulating many-body quantum systems.  We combine hybrid quantum variational techniques with the recent development of \textit{programmable} analog quantum simulators, which provide us with potentially \textit{scalable} analog quantum devices, capable of performing  a restricted set of high-fidelity quantum operations.  Essentially,  VQS provides a framework for quantum simulation of complex  lattice models beyond AQS and DQS, producing \textit{best} answers for \textit{given} quantum resources, as available in today's laboratories.

\begin{figure}
 \includegraphics[width=\columnwidth]{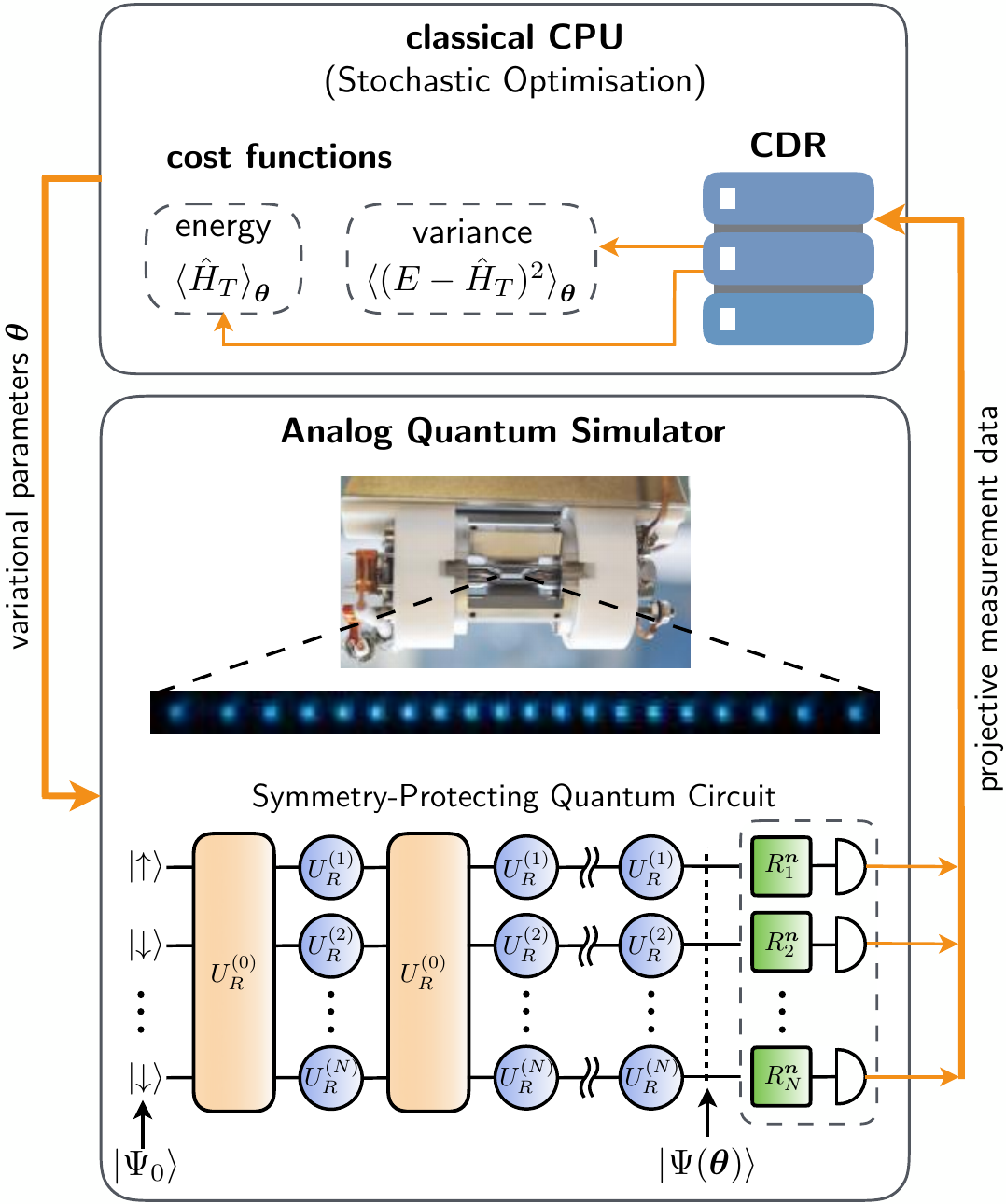}
 \caption{ \label{fig:VQS}
 {Classical-Quantum Feedback-Loop of VQS:}
Optimisation of cost functions with function evaluations performed on a programmable 20-qubit trapped-ion analog quantum simulator.
   Variational control parameters are passed to the
analog quantum simulator, which generates trial states as quench dynamics from a set of resource Hamiltonians with symmetry-protecting quantum circuits, consisting of entangling and single qubit operations, corresponding to quantum resources. Measurement results for correlation functions are obtained by rotating the qubits to the proper basis (green boxes) followed by quantum projective measurements in the standard basis. A central data repository (CDR) stores the information on the obtained many-body correlation functions and allows for data re-evaluation for different Hamiltonian parameters (see Appendix \ref{App:CDR}).}
\end{figure}

Quantum variational techniques have been developed in quantum chemistry and in the context of classical optimisation~\cite{VQAMcClean2016,MollGambetta2018,Farhi2014,OMalley2016}. Here a quantum device prepares a family of (possibly) highly entangled variational trial states $|\Psi(\paramvec)\rangle$ with $\paramvec$ a vector parameterising the quantum circuit which generates the quantum state. 
In a closed feedback loop (Fig.~\ref{fig:VQS}), a classical computer then optimises the parameters $\paramvec$ over a relevant cost function: for example, to compute ground states  of a given target Hamiltonian $\hat{H}_T$ we seek to minimise $\bra{\Psi(\paramvec)} \hat{H}_T \ket{\Psi(\paramvec)}$. The role of the quantum device is to perform the classically difficult
task of evaluating observables from highly entangled states. For lattice spin models, a generic Hamiltonian, $\hat{H}_T = \sum_i \hat{h}_i$,
is decomposed into strings $\hat{h}_i \propto  \hat{\sigma}^{a_1}_{j_1} \ldots  \hat{\sigma}^{a_s}_{j_s}$ of Pauli operators $\hat{\sigma}^{a}_{j}$, where $j$ labels the lattice site and the $a$ Cartesian coordinates.
The energy cost function then becomes a sum of expectation values of correlation functions $\bra{\Psi(\paramvec)}\hat{h}_i \ket{\Psi(\paramvec)}$, which are individually measured on the quantum device. In this procedure, the target Hamiltonian $\hat{H}_T $ only exists as a set of measurement prescriptions and never needs to be realised physically in the laboratory, endowing the experimenter with great freedom in the choice of models to quantum simulate. In addition, once the classical parameters $\paramvec$ corresponding to (a good approximation of) the ground state are found, the state can be re-prepared at will and is available for further study.

Small-scale, pioneering experiments with hybrid algorithms running on quantum computers of 2 to 6 qubits have been shown to successfully tackle problems in quantum chemistry
\cite{Hempel_2018,Otterbach_2017,OMalley2016,Peruzzo2014,Dumitrescu_2018}, and more recently in condensed matter and high-energy physics ~\cite{Kandala2017,Klco_2018}. 
In these experiments, scaling to larger system sizes is limited by available quantum computing hardware, as well as a rapid increase in
the number of variational parameters $\paramvec$. Moreover, an inherent challenge of quantum variational techniques is the rapidly increasing number of measurements needed for the classical computer to successfully navigate the resulting high-dimensional energy landscape with increasing system size, from which information can only be obtained through inherently noisy projective measurements. Finally,
when variational techniques can be scaled up to a regime where classical simulations become intractable, we are faced with the long-standing
problem of verification of the answers produced,
i.e., assessing how close the variational trial states and energies are to the exact values.
To date, scalability and applicability of VQS remain key challenges \cite{VQATroyer2015,VQSPontryagin2017,QAOAPichler2018,Carleo2017,LloydMontangero2014}.

Here we demonstrate how, by using existing AQS setups for VQS and focusing on lattice models, we can scale VQS up to 
20 qubits, the maximum available in our setup. In addition, we show quantitative
self-verification of the acquired results by measuring {\it algorithmic} error bars of the final energies, i.e.~the uncertainty on the approximate ground state energy resulting from the variational ansatz state at finite circuit depth. Such algorithmic error bars are
evaluated on the quantum device, by directly measuring the variance of the target Hamiltonian. The key elements behind this advance are: (i) the use of a programmable analog quantum simulator as a potentially scalable, although non-universal, quantum hardware; (ii) our focus on quantum lattice models of condensed matter and high-energy physics, and incorporating intrinsic symmetries in trial states of VQS, which allow us to reduce significantly the number of variational parameters to be optimised; (iii) an advanced global optimisation algorithm specifically suited for noisy, high-dimensional and gradient-free optimisation problems, and a reuse of measurement data to efficiently find ground states of whole classes of Hamiltonians.
Combining these elements, we demonstrate below VQS of the lattice Schwinger model~\cite{Schwinger_62,HAMER1982413} with up to 20 qubits on a programmable trapped-ion quantum simulator~\cite{ProbeRenyi2018} that naturally implements a long-range transverse XY spin model~\cite{PorrasCiracGateExperiment,MonroePropagation2014,Jurcevic2014} and single site spin rotations, as quantum resource.

\section{Variational quantum simulation of the Schwinger model with trapped-ions}
\label{sec:targetresource}

In VQS there exists a clear distinction between the \textit{target} model Hamiltonian to be studied, and the \textit{resources} available in the laboratory that are used to produce trial states $\ket{\Psi(\paramvec)}$. We now describe in turn the Schwinger model as our target problem, and list available quantum resources provided by the trapped-ion analog quantum simulator.

{\it Target --  lattice Schwinger model:} The Schwinger model on a lattice is a paradigmatic formulation of 1D quantum electrodynamics, and a prototype of an Abelian lattice gauge theory \cite{SchwingerModel1951}.
It describes the interactions between a scalar fermion field, representing both matter and antimatter with electric charges, and an Abelian $U(1)$ gauge field as a quantised electromagnetic field. We use a Kogut-Susskind encoding to map fermionic configurations to a spin-$1/2$ lattice, where a spin down (resp.~up) on an odd (even) lattice site indicates the presence of a positron (electron) (see Appendix \ref{sec:elimination} for details). For open boundary conditions, and with a Jordan-Wigner transformation \cite{Encoding_97}, our target Hamiltonian reads
\begin{align}\label{eq:schwinger_spin}
\Htarg \!= w
\!\! \sum_{j=1}^{N-1} \! \left[ \hat{\sigma}_j^+  \! \hat{\sigma}_{j+1}^- \! \!+ \! \text{H.c.}  \right] \! + \!
 \frac{m}{2}\!\sum_{j=1}^N (-1)^j  \hat{\sigma}_j^z 
\!+ \! \bar{g} \!\sum_{j=1}^{N}\! \hat{L}_j^2,
\end{align}
where $j$ labels the lattice site, for a system of length $N$, and $\hat{\sigma}^a_j$ are Pauli operators.
Here the first term describes the creation or annihilation of a particle-antiparticle pair, mapped to a spin flip-flop term with coupling $w$. In the following, we set $w=1$ as the energy unit. The second term is the matter mass term, with bare mass $m$.
The last term, with coupling $\bar{g}$, is the electric field energy.
The 1D character of our model allows the electric field $\hat{L}_j$ to be eliminated due to Gauss' law \cite{HAMER1982413},
\begin{equation}
 \hat{L}_{j} - \hat{L}_{j-1} = \frac{1}{2} ( \hat{\sigma}^z_j + (-1)^j), 
\end{equation}
i.e.~we express the electric field in terms of Pauli operators, $\hat{L}_j = \epsilon_0 - \frac{1}{2} \sum_{\ell = 1}^{j} ( \hat{\sigma}^z_\ell + (-1)^\ell)$, in $\Htarg$. Here $\epsilon_0$ is a background electric field, which is set to zero.
The Hamiltonian $\Htarg$ thus reduces to an effective spin-$1/2$ model with exotic long-range interactions originating from squaring $\hat{L}_j$. We note that $\Htarg$ exhibits a charge conservation symmetry $ \hat{\sigma}^z_\text{tot} = \sum_{j}  \hat{\sigma}^z_j$, and, within the charge symmetry sector $ \hat{\sigma}^z_\text{tot} = 0$, a $\CPsym$ symmetry. Such a symmetry consists of charge conjugation and spatial reflection, which are not separately present due to the staggered fermion encoding, at finite size and with open boundary conditions (see Appendix \ref{sec:elimination}).

The complexity of the lattice Schwinger Hamiltonian has so far prevented a direct laboratory implementation using AQS \cite{Preskill2012,Klco_2018}. DQS of the lattice Schwinger model  has been reported in \cite{Martinez2016} with four qubits and four Trotter steps requiring a total of 220 quantum gates. 
In contrast, we show below that VQS allows well-converged quantum simulation of ground states of the lattice Schwinger model, for up to $20$ qubits, and with the unique possibility to self-verify the answers on the quantum device itself.

{\it Quantum resource -- trapped-ion analog quantum simulator:}
Our experimental simulator consists of $N = 20$ trapped $^{40}\text{Ca}^+$ ions that are confined as a 1D string in a linear Paul trap. Each ion represents a qubit, or spin-$1/2$, encoded in the internal electronic levels $\ket{\downarrow} = \ket{S_{1/2},m_j=1/2}$, $\ket{\uparrow} = \ket{D_{5/2},m_j=5/2}$
which are used as the computational basis \cite{Jurcevic2014}. A bichromatic laser beam, off-resonantly coupling all the ions to the transverse phonon modes, realises an XY Hamiltonian
\begin{equation} \label{eq:ionIsing}
\Hres^{(0)} = 
 \sum_{i=1}^{N-1} \sum_{j=i+1}^N J_{ij} (\hat{\sigma}_i^+  \hat{\sigma}_j^- +  \hat{\sigma}_i^-  \hat{\sigma}_j^+) + B \sum_{i=1}^N \hat{\sigma}_i^z,
\end{equation}
where $J_{ij} \simeq J_0 /|i-j|^{\alpha}$, with $0 < \alpha < 3$ (see Appendix \ref{App:ExperimentalDetails}), are long-range antiferromagnetic couplings, and $B$ is an effective, uniform magnetic field
\cite{PorrasCiracGateExperiment}. This Hamiltonian is capable of creating highly entangled states \cite{MonroePropagation2014,Jurcevic2014,ProbeRenyi2018}
and constitutes our first resource for building variational states. It is complemented by Hamiltonians $\hat{H}_R^{(j)}= \frac{\Delta_0}{2}\hat{\sigma}_j^z$ generating local spin rotations around the $z$ axis, which are realised by a steerable, strongly focused off-resonant laser beam inducing a level shift $\Delta_0$.

The set of Hamiltonians $\{ \Hres^{(j)} \}$ provides us with a quantum resource to generate the family of trial states 
$|\Psi({\paramvec})\rangle = \exp(- i \theta_k \Hres^{(i_k)}) \cdots \exp(- i \theta_1 \Hres^{(i_1)}) |\Psi_0\rangle$. Here the label $i_m$ selects the resource Hamiltonian $\Hres^{(i_m)}$ employed in the $m$-th operation,
$\ket{\Psi_0}$ is a simple (e.g.~product) initial state, and $\paramvec = (\theta_1, \ldots , \theta_k)$ is a parameter vector. Our variational trial states are generated by first preparing the ions in one of the N\'eel-ordered states,
e.g.~$\ket{\Psi_0} =\ket{\uparrow \downarrow \cdots \uparrow \downarrow}$ representing the bare vacuum, i.e.~the ground state for $m\rightarrow \infty$. We subsequently subject $\ket{\Psi_0}$ to a quantum circuit composed of alternating {\it layers} of unitary operations: the odd layers are each composed of an entangling operation $U_R^{(0)}(\theta) =\exp ( {-i\theta\hat{H}_R^{(0)}} )$, while the even layers are each a set of local light shift operations $U_R^{(j)}(\theta)=\exp ({-i
\theta\hat{H}_R^{(j)}})$ acting on every site $j$
(see Fig.~\ref{fig:VQS}). The unitaries used for preparing the trial state $|\Psi({\paramvec})\rangle$ from $|\Psi_0\rangle$ keep the state in a decoherence-free subspace with respect to the major experimental sources of decoherence \cite{ProbeRenyi2018}, thereby achieving a high-fidelity state preparation. 

The resource Hamiltonians match the symmetries of the target problem well.
In fact, $\hat{H}_R^{(0)}$ preserves both $\hat{\sigma}^z_{\text{tot}}$ and $\CPsym$ symmetries on the level of quantum hardware. The single-qubit operations can be enforced to be symmetry-protecting
by constraining the local $z$-rotations, in the same layer, acting on sites $j$ and ${N+1-j}$, to $\theta^{(j)}=-\theta^{(N+1-j)}$.
This reduces the number of variational parameters of each single-qubit layer
to $N/2$ (Fig.~\ref{fig:VQS}), while restricting the global search to the symmetry sector $\mathcal{S}$ of interest ($ \hat{\sigma}^z_{\text{tot}} \ket{\Psi} = 0$, $\CPsym \ket{\Psi}=+ \ket{\Psi}$).
Furthermore, as the number of lattice sites increases, we expect the model to exhibit approximate translational symmetry within a bulk-region $\mathcal{B}$ in the centre of the lattice. This can be exploited, in order to further reduce the number of parameters, by enforcing 
$\theta^{(j)} = \theta^{(j+2)}$ for all $\{j, j+2\} \in \mathcal{B}$, i.e.~far from the edges.

Finally, projective measurements - yielding spin correlation functions, as required for the evaluation of $\langle \Htarg\rangle_{\mathbf \theta}$ and $\langle \Htarg^2\rangle_{\mathbf \theta}$ - are achieved by local spin rotations prior to state detection in the computational basis. To determine $\langle \Htarg\rangle_{\mathbf \theta}$,  measurements in 3 different bases are performed, while for  $\langle \Htarg^2\rangle_{\mathbf \theta}$  we require $3 N$ different bases (see Appendix \ref{App:MS}). 

\begin{figure*}
 \includegraphics[width=0.95\textwidth]{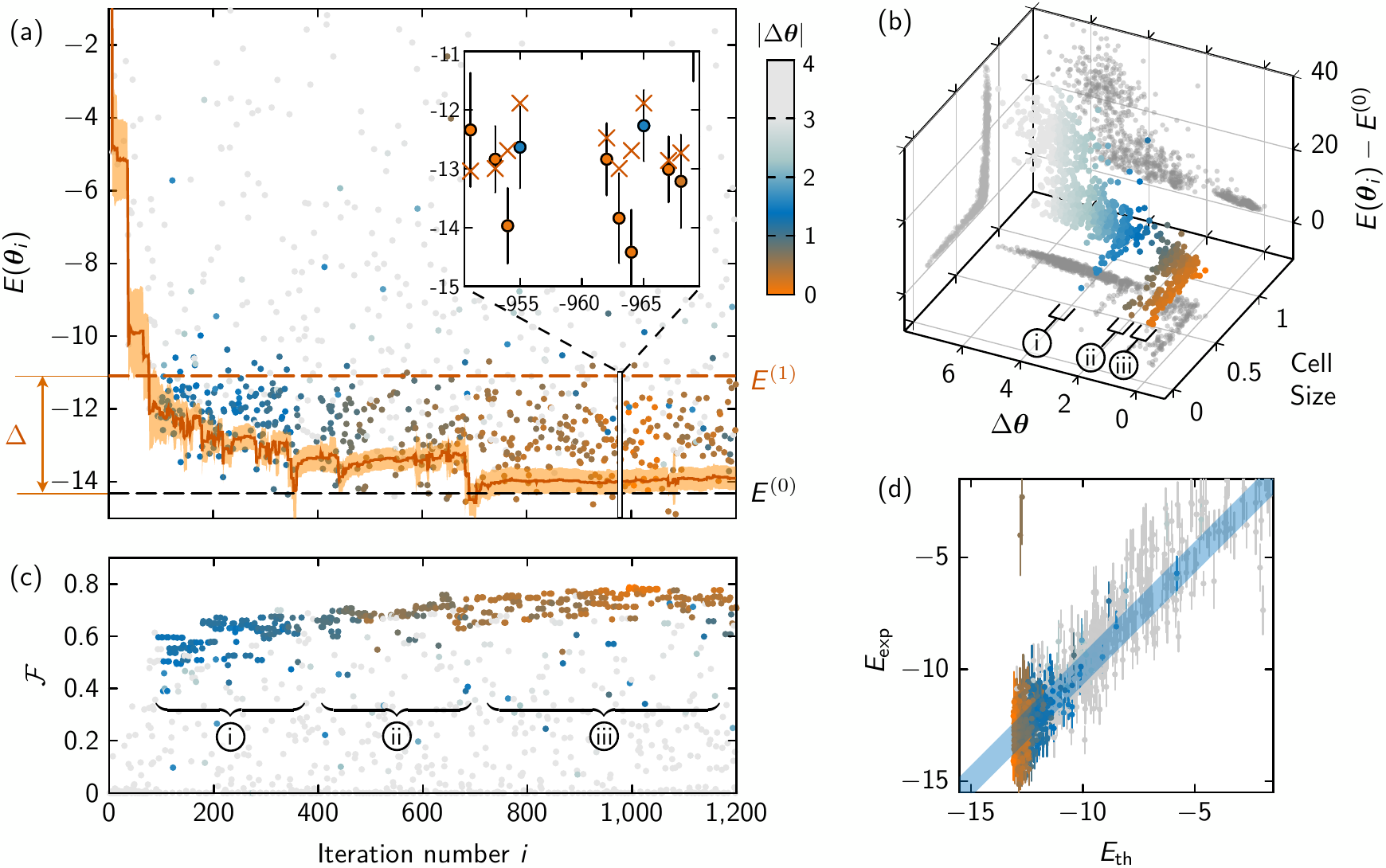}
 \caption{
{Schwinger ground state (20 ions).} {(a) Optimisation trajectory:} Convergence of experimental
energies $E(\paramvec_i)\equiv\bra{\Psi(\paramvec_i)}\Htarg\ket{\Psi(\paramvec_i)}$ (dots)
vs.~iteration number $i$ of the DIRECT optimisation algorithm (see text), for  $m\!=\!0.9,w\!=\!\bar{g}\!=\!1$. To ensure global convergence, the search algorithm not only attempts to refine the current minimum, but also keeps exploring new parts of the parameter space, leading (often) to high energy values, even in the later stages of the run.
Energy values $E({\paramvec_i})$  are colour-coded to indicate the Euclidian distance of $\paramvec_i$ to the final optimised parameter vector $\paramvec_\text{opt}$, as selected by theoretical fidelity (see panel (c)). The solid red line indicates the algorithm's current estimate of the groundstate energy and its $2\sigma$ uncertainty (shaded area), from modelling the thus far observed energies as jointly gaussian distributed random variables (see Appendix \ref{App:OA}). Inset: Close-up of a late stage of the optimisation, where statistical errorbars (as defined in Appendix \ref{App:VQSworkflow}) are displayed, and theoretically simulated values are plotted as crosses. 
{(b) Visualisation of sampled energy landscape}: Sampled energies $E(\paramvec_i)$ relative to the ground state energy, plotted versus the parameter distance $\Delta\paramvec$ and the size of the domain in parameter space (cell) that each point represents internally in the optimisation algorithm (see Appendix \ref{App:OA}). Distinct local minima are visible as 'fingers', marked (i), (ii) and (iii), extending towards smaller cell sizes, indicative of an increasingly fine sampling of the parameter space near a local minimum.
{(c) Fidelity of many-body wavefunction:} Theoretical
fidelities $\mathcal{F}$ computed for $\paramvec_i$, with a rough indication of the iterations during which each of the minima (i), (ii), (iii) of panel (b) provided the current best solution.
{(d) Correlation experiment-theory}: Experimentally measured energy ($E_\text{exp}$) versus numerically simulated ($E_\text{th}$) energies, showing agreement within $2 \sigma$ (blue shaded area).
}
\label{fig:20ions}
\end{figure*}


\section{Results}
\label{sec:results}

We now present experimental results for ground state determination, self-verification, and quantum phase transitions for the lattice Schwinger model, within the VQS framework, using the trapped-ion analog quantum simulator as the experimental platform.
Variational quantum many-body states were built on setups with up to 20 ions, up to $6$ layers of circuit depth, and up to $15$ variational parameters. We compare our experimental results with theoretical expectations from simulating optimisation runs on a classical computer, assuming ideal coherent dynamics generated by the resource Hamiltonians. 

Global parameter optimisation was performed with a variant of the dividing rectangles (DIRECT) algorithm \cite{JonesOriginalDirect93,LiuDirect2015,NicholasDirect2014}, that divides the parameter space into regions called \textit{cells}, which are represented by a single cost function evaluation. Cells are subdivided when deemed sufficiently promising to harbour the global minimum, taking into account cost function values and cell sizes (see Appendix \ref{App:OA}). To each optimisation run, we assign a total budget of (up to) $10^5$ calls to the quantum simulator, where a single call involves variational state preparation and projective measurement of the qubits in a given basis. To determine the energy expectation value for given parameters, initially $30$ projective measurements per basis are performed. Additional samples are taken when refinement is needed, e.g. when two similarly sized cells are competing candidates for a subdivision step, but it is unclear which function value is lower due to the statistical uncertainty resulting from a finite number of projective measurements.

{\it Variational Schwinger ground state for 20 ions --} 
An optimisation trajectory consisting of energy evaluations $E({\paramvec_i})\equiv \bra{\Psi(\paramvec_i)}\Htarg\ket{\Psi(\paramvec_i)}$ for 20 ions is shown in Fig.~\ref{fig:20ions} (a), as a function of the iteration number $i$ of the optimisation algorithm. Energies and wavefunctions were found to converge with a circuit of depth 6. We approximate translational invariance in a bulk region of $14$ central sites, resulting in a variational scheme with $15$ parameters.
For reference, we plot the exact ground state energy $E^{(0)}$ and first excited state $E^{(1)}$ in the zero-magnetisation sector (corresponding to an intra-sector energy gap $\Delta \!= \!E^{(1)}\! -\! E^{(0)}$), obtained from exact diagonalisation (ED) of the lattice Schwinger model.
Our global search algorithm continuously explores new parts of the parameter space, leading to a large spread in the sampled energies. 
The algorithm maintains an internal model of the energy landscape (see Appendix \ref{App:OA}), that is continuously updated as more experimental data is gathered (Fig.~\ref{fig:20ions} (a), red, solid line).
Fig.~\ref{fig:20ions} (b) visualises the sampled energy landscape. An initial local minimum is visible (i), but surpassed by more promising regions (ii) and (iii) that ultimately delivered the final best answer.
Fig.~\ref{fig:20ions} (c) shows theoretically computed fidelities, defined as $\mathcal{F} = |\braket{\Psi_G|\Psi^{\text{Sim}}({\paramvec_i})}|^2$ between the exact ground state $\ket{\Psi_G}$ and variational wave function $\ket{\Psi^{\text{Sim}}({\paramvec_i})}$ simulated numerically with parameters $\paramvec_i$ taken from the \textit{experimental} run. The final fidelity for the quantum many-body ground state with 20 ions approaches $0.8$. A similar optimisation run for 16 ions found multiple local minima and reached a final fidelity approaching $0.9$ (Appendix \ref{App:Fig} Fig. \ref{fig:16ions}). The increase in fidelity when lowering the number of ions is attributed to less decoherence and a better initial state preparation fidelity.
Fig.~\ref{fig:20ions} (d) shows agreement, in terms of energies,  between experimental results and theoretical simulations for corresponding trial states $\ket{\Psi(\paramvec_i)}$. Typical statistical error bars are drawn as a blue band, of thickness $2 \sigma$.

{\it Variational Schwinger ground state and self-verification for 8  ions --} For a more in-depth analysis of the VQS method, we turn to a smaller system size of 8 ions.
\begin{figure}
 \includegraphics[width=\columnwidth]{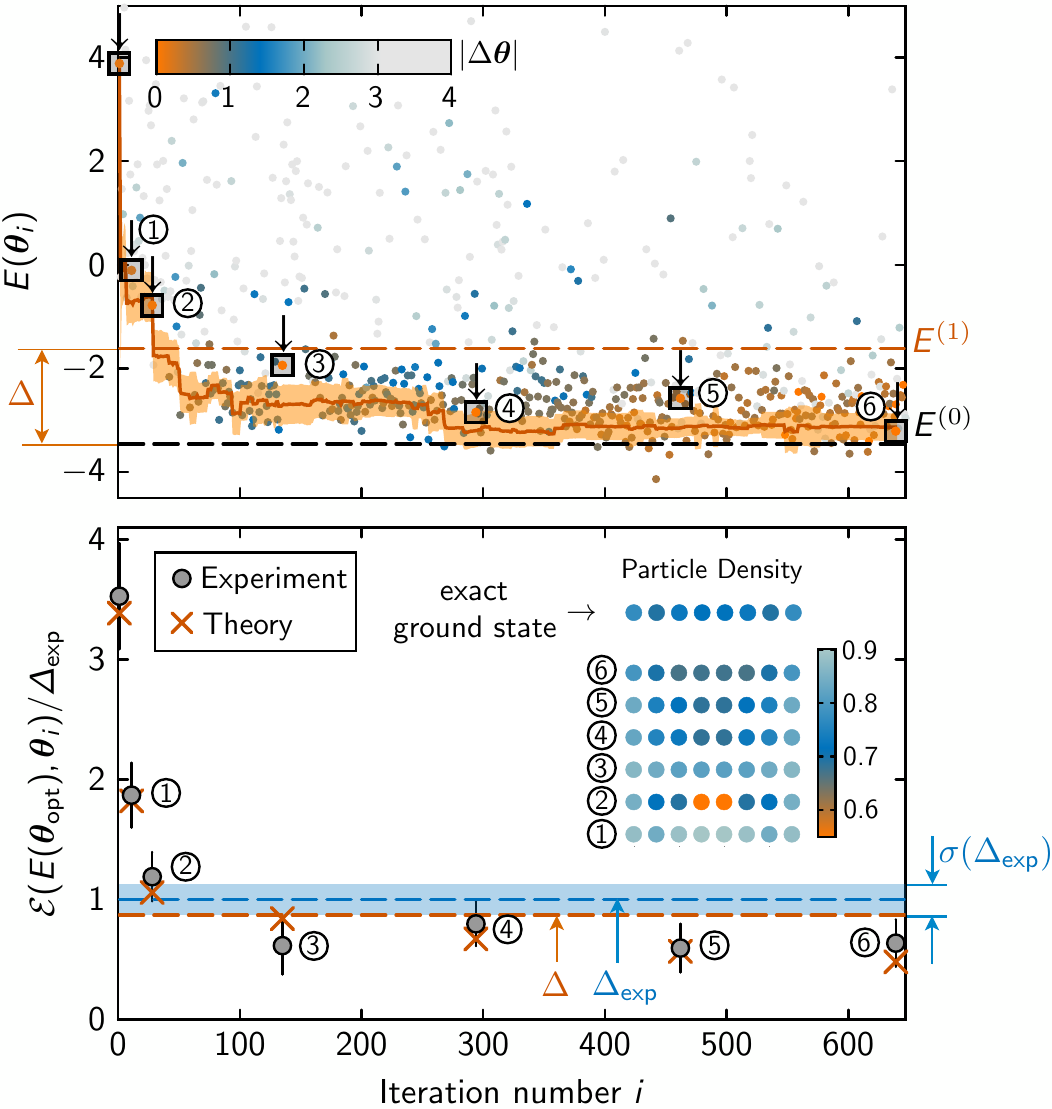}
 \caption{ \label{fig:verification}
{Schwinger ground state (8 ions).} {(top panel) Optimisation trajectories:} Convergence of experimental
energies $E({\paramvec})\equiv\braket{\Htarg}_{\paramvec}$
for 8 ions ($m\!=\!0.1,w\!=\!\bar{g}\!=\!1$) vs.~number of iterations of the global optimisation algorithm (compare Fig.~\ref{fig:20ions}). 
{(bottom panel) Self-verification of VQS:}
Measurement of algorithmic
error $\mathcal{E}$, in units of the experimentally determined energy gap $\Delta_\text{exp}$,
vs.~iterations, for
data points indicated by squares in (a), showing
numerical simulation (crosses) and experimental data (filled circles), including statistical error bars (see Appendix \ref{App:VQSworkflow}). For comparison, the exact energy gap $\Delta$ is also indicated.
The inset shows the convergence of the site-resolved particle density $\braket{\hat{n}_j}$ for data points 1-6. }
\end{figure}
In Fig.~\ref{fig:verification} (upper panel) we show an optimisation run for the Schwinger ground state, using 4 layers of circuit depth and no bulk enforcement, resulting in $10$ variational parameters.
The experimental run provides us with a final energy of $E({\paramvec}_\text{opt})= -3.24 \pm 0.36$, corresponding to a fidelity approaching $0.95$. This is equivalent to an excitation $( E({\paramvec}_\text{opt}) - E^{(0)} )$ of $(11 \pm 18) \%$
of the energy gap $\Delta_\text{exp}$, where  $\Delta_\text{exp} = 2.11 \pm 0.24$ itself is experimentally evaluated via a quantum subspace expansion strategy (see Appendix \ref{App:QSE}) \cite{Colless_2018}.

Our VQS scheme has the built-in feature of self-verifying its quantum simulation results.
 For an approximate ground state energy $E({\paramvec})\equiv\braket{\Htarg}_{\paramvec}$, the algorithmic error bar of VQS (as a function of the depth of the circuit) is defined by measuring the Hamiltonian variance $\mbox{$\mathcal{E}^2(E, {\paramvec})=\langle (\Htarg - E)^{2} \rangle_{{\paramvec}} $}$.
 For an exact eigenstate, the Hamiltonian variance $\mathcal{E}^2(E(\paramvec), \paramvec)$ is zero, whilst for an approximate wave function $\ket{\Psi({\paramvec})}$,
it provides an uncertainty estimate, according to the bound $|E_\ell - E({\paramvec})|\leq \mathcal{E}(E({\paramvec}), {\paramvec})$, with $E_\ell$ the exact eigenstate energy closest to $E({\paramvec})$. The algorithmic error is to be contrasted with the {\it statistical} error resulting from predominantly shot noise, which is responsible for the noisy energy landscape (see Appendix \ref{App:OA}).
Self-verification of the ground state energy is illustrated in Fig.~\ref{fig:verification} (lower panel). Here we plot the measured algorithmic error $\mathcal{E}$, for selected trial states, marked by the squares in Fig.~\ref{fig:verification} (upper panel). The plot demonstrates convergence to an energy eigenstate, manifested by the asymptotically decreasing algorithmic error with final value $\mathcal{E}/\Delta_\text{exp} = 0.64\pm 0.20$. The variances $\mathcal{E}^2$
are determined from projective measurements in 24 different bases (see Appendix \ref{App:MS}). As the circuit depth increases, the lowest achievable algorithmic error decreases (Appendix \ref{App:Fig}, Fig. \ref{fig:8IonsAlgoError}). To illustrate convergence of observables associated with the variational wave function, we plot as inset the particle density $\braket{\hat{n}_j} \equiv \frac{1}{2} (1 + (-1)^j\braket{ \hat{\sigma}^z_j})$ at lattice site $j$ obtained from the experiment for the enumerated points marked by 1 to 6, and compare them with the exact ground state result.

\begin{figure}
 \includegraphics[width=\columnwidth]{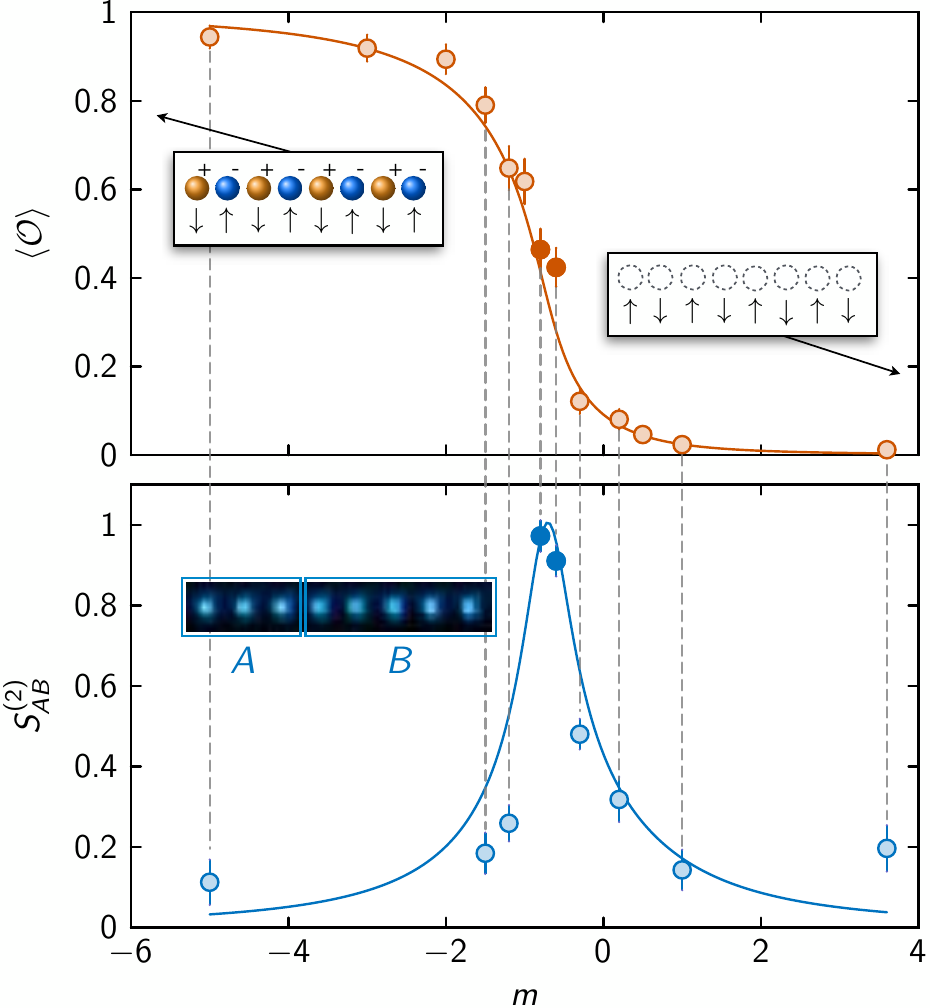}
 \caption{ \label{fig:qpt}
 {Quantum phase transition:}
The top panel shows the order parameter $\braket{\hat{\mathcal{O}}}$ defined in the  main text, and the lower panel displays the second-order R\'enyi entanglement entropy $S^{(2)}_A$ (with A-B bipartition shown in inset)  as a function of mass $m$ from negative to positive values, with $m_c \sim -0.7$ being the critical point ($w\!=\!\bar{g}\!=\!1$). Measured data are represented by circles, and solid lines are exact theoretical results from ED. For the light orange circles, variational trial states are prepared by applying quantum circuits of depth 4 (10 parameters).
The initial starting state $\ket{\Psi_0}$ for VQS are $\ket{\downarrow \uparrow \ldots \downarrow \uparrow}$ and  $\ket{ \uparrow \downarrow \ldots \uparrow \downarrow }$, for $m<m_c$ and $m>m_c$ respectively.
For the two points in proximity of the phase transition (filled blue circles) we use a circuit of depth $5$, adding an additional entangling layer (11 parameters),
to achieve convergence.}
\end{figure}

{\it Quantum phase transition for 8 ions --} VQS allows the study of quantum phase transitions by monitoring the change in the ground state as a function of a Hamiltonian parameter.  We illustrate this in Fig.~\ref{fig:qpt}  by varying the bare mass $m$ in the lattice Schwinger model from positive to negative values.
Under these conditions, the Schwinger model is known to undergo a second-order phase transition, belonging to the universality class of the 1D Ising model in a transverse field (notice that the resource Hamiltonian exhibits no such quantum phase transition). In the Schwinger model the two limits of $m\rightarrow \pm \infty$ correspond to the bare vacuum, and proliferation of electron-positron pairs, respectively (see insets in Fig.~\ref{fig:qpt} (upper panel)). The critical point for zero background field $\epsilon_0 = 0$ occurs at negative bare mass $m_c \sim -0.7$ \cite{Coleman1976239,Byrnes_2002}. We experimentally detect the phase transition by measuring the order parameter $\braket{\hat{\mathcal{O}}}= \frac{1}{2N(N-1) }\sum_{i, j> i} 
 \langle (1+(-1)^i \hat{\sigma}^z_i) (1+(-1)^{j} \hat{\sigma}^z_{j}) \rangle$
for ground states optimised via VQS, and compare them with exact theoretical results from ED (orange solid line).
The optimisation was accelerated by re-evaluating previously sampled data for different values of $m$ (see Appendix \ref{App:CDR}).
To quantify entanglement across the phase transition we measure, in Fig.~\ref{fig:qpt} (lower panel),  the second-order R\'enyi entropy \cite{ProbeRenyi2018}, $S_A^{(2)}= - \text{Tr} [ \log_2 (\rho_A ^2) ]$, where $\rho_A$ is the reduced density matrix in the bipartition A-B shown in the inset of Fig.~\ref{fig:qpt} (lower panel). 
Convergence at the critical point is achieved with a quantum circuit of 5 layers (blue filled circles).


\section{Summary and Outlook}
\label{sec:outlook}

We have experimentally demonstrated  self-verifying VQS of the lattice Schwinger model  with a programmable trapped-ion analog quantum simulator, implemented as a classical-quantum feedback loop. We have considered system sizes of up to 20 qubits, and optimised up to 15 parameters globally using a variation of the dividing rectangles (DIRECT) algorithm within a fixed budget of up to $10^5$ calls to the quantum co-processor.  Our algorithms and techniques apply immediately to a broad class of lattice models in condensed matter and high-energy physics, as well as programmable quantum simulators built with other platforms \cite{Bernien51atoms,BrowaeysQSim2016,Houck2012}. A key element in VQS of lattice models is incorporating symmetries in quantum circuits to reduce the number of variational parameters, strengthening the potential scalability of hybrid simulations \cite{VQATroyer2015,VQSPontryagin2017,QAOAPichler2018,Carleo2017,LloydMontangero2014,Doria2011}.
While we have studied a lattice Schwinger model  with open boundary conditions, our computations for large system sizes benefit greatly from the (approximate) translational invariance of the bulk. This provides an interesting perspective on scaling VQS to large and, in particular, infinite systems to be simulated with comparatively few variational parameters, as a quantum analog of iMPS and iPEPS \cite{iMPS,iPEPS}. 
Finally, the present work can be extended to 2D and 3D analog quantum simulators \cite{Bollinger2DQsim2012,WeimerQSim2010}, and fermionic Hubbard systems \cite{GreinerQSimFermion2018,GrossQSim2017,EsslingerQSimFermion2010}. This VQS toolbox opens interesting perspectives to address longstanding equilibrium problems in condensed matter and high-energy physics, that are inaccessible for Markov Chain Monte Carlo methods due to sign problems, including models with finite baryon density and so-called topological terms.

\vspace{1ex}
\noindent${\ast}$ CK, CM and RvB contributed equally to this work.

\vspace{1ex}
\noindent\textit{Acknowledgments --} The authors thank J.~Bollinger, A.~Elben, K.~Holmstr\"om, K.~Jansen, M. ~Lukin, E.~A.~Martinez, S.~Montangero, P.~Schindler and U.~J.~Wiese for discussions. The numerical results presented have been achieved (in part) using the HPC infrastructure LEO of the University of Innsbruck.
The research at Innsbruck is supported by the ERC Synergy Grant UQUAM, by the European Research Council (ERC) under the European Union's Horizon 2020 research and innovation programme under grant agreement No 741541, the SFB FoQuS (FWF Project No. F4016-N23), QTFLAG - QuantERA and the Quantum Flagship PASQUANS. We acknowledge A.~Elben and B.~Vermersch for development of the software for evaluating the R\'enyi entropies measurement.

\appendix

 \section{Global search algorithm with noisy cost function}
\label{App:OA}
The variational optimisation algorithm running on the classical computer is tasked with finding the global minimum of a high-dimensional energy landscape, in the space of parameters $\paramvec$, that only reveals itself through inherently noisy measurements of quantum observables, while typically featuring multiple local minima. 
We employ a modified version of the DIviding RECTangles algorithm (DIRECT) \cite{JonesOriginalDirect93, DirectConvergence2004, NicholasDirect2014, LiuDirect2015}, which divides the search space into so-called hypercells, where each cell is represented by a single cost function evaluation taken in its interior.
Promising cells are sampled at a finer scale by subdividing them into smaller cells, prioritising cells with low energy values, as well as cells with large sizes. The cell size is a relevant quantity in the algorithm's decision making, since larger cells have more unexplored territory, and are hence statistically more likely to harbour the global minimum, compared to smaller ones. The DIRECT algorithm attempts to strike a balance between {\it exploitation}, i.e. refining an already good point, and {\it exploration}, filling in the unknown areas in parameter space. The exploration stages ensure that the algorithm is guaranteed to ultimately find the global minimum. We found that the DIRECT algorithm solves the optimisation problem of VQS very efficiently for dimensions up to 20. For higher dimensions, further modifications of the algorithm are likely required \cite{Tavassoli2014}.

During the optimisation, the algorithm maintains an internal representation, called \textit{metamodel}, of the energy landscape in the form of a Gaussian Process \cite{RasmussenWilliams}, modeling the data as jointly distributed gaussian variables with a squared exponential covariance kernel, $2\pi$-periodic in the parameters associated with the single particle rotations. The metamodel's hyperparameters determining the correlation lengths are obtained from a maximum likelihood estimation. The internal model yields not only predictions for the mean values of the energy landscape, but also provides values for the inherent uncertainty of a prediction.
The metamodel is used in a greedy step in the algorithm, sampling at the predicted global minimum according to the model, as well as aiding in the selection of points for subdividing hypercells \cite{LiuDirect2015}. Predictions for the global minimum are shown in Figs. \ref{fig:20ions}(a) and \ref{fig:verification}(upper panel) as solid lines, with the shaded areas indicating the $2\sigma$ confidence interval of the prediction.
 
Measurements of cost functions are affected by statistical errors \cite{VQSError2017}. In our case, the errors are introduced due to a finite number of projective measurements dedicated to the energy evaluation (shot noise), temporal fluctuations in the experimental controls (control noise),
and infidelities in the initial N\'eel state preparation.
Such intrinsic noise must be taken into account by the search algorithm in order to achieve robust convergence.
An additional {\it refinement} stage is therefore added to the algorithm, where selected points in parameter space are once again passed to the analog quantum simulator and the cost function is re-evaluated. The additional measurements are then combined with the previous ones, to reduce statistical error attached to this landscape point. 

The algorithm is provided with a finite, total measurement budget which it can spend during the optimisation process. In order not to waste this budget on high energy points, we initially invest only a relatively low number of measurements at each unexplored point. The number of initial samples can be as low as 6 per measurement basis for the 8-ion problem, although due to technical overheads we maintain a minimum of 30 per measurement basis. The low number of initial samples typically suffices to provide a rough estimation of the energy on which the algorithm can base its subsequent decisions, i.e. which hypercells to subdivide. If the variance of the cost function measurements is too large, the algorithm can request refinement steps at datapoints already sampled in order to increase the probability of correctly deciding which cells to subdivide.
The algorithm selects when to perform the refinement steps and how many measurements to spend in this stage, based on methods from decision theory and Optimal Computing Budget Allocation (OCBA) \cite{OCBA2008, NicholasDirect2014}.

In the present work, the variational algorithm continues requesting measurements from the quantum device until exhaustion of the allocated resource budget. Alternative stopping criteria could involve a comparison against a fixed goal, e.g., setting {\it a priori} an energy precision threshold $\mathcal{E}_\text{T}$. The algorithm would then be terminated as soon as the algorithmic error bar $\mathcal{E}(E(\paramvec), \paramvec)$ (plus its own statistical error bar) becomes smaller than $\mathcal{E}_\text{T}$. In practice, however, this requires the measurement of the target Hamiltonian variance during the optimisation, which would consume a part of the measurement budget that would be no longer available for direct optimisation. Another typical stopping criterion for optimisation algorithms involves an evaluation of stationarity, i.e. if the proposed solution is no longer improving as more iterations are added. For our optimisation algorithm, such a comparison could for instance be made against the predicted minimum of the metamodel.

 \section{Protecting symmetries of the Schwinger model}
\label{App:ProtectingSymmetries}
\begin{figure*}
 \includegraphics[width=0.95\textwidth]{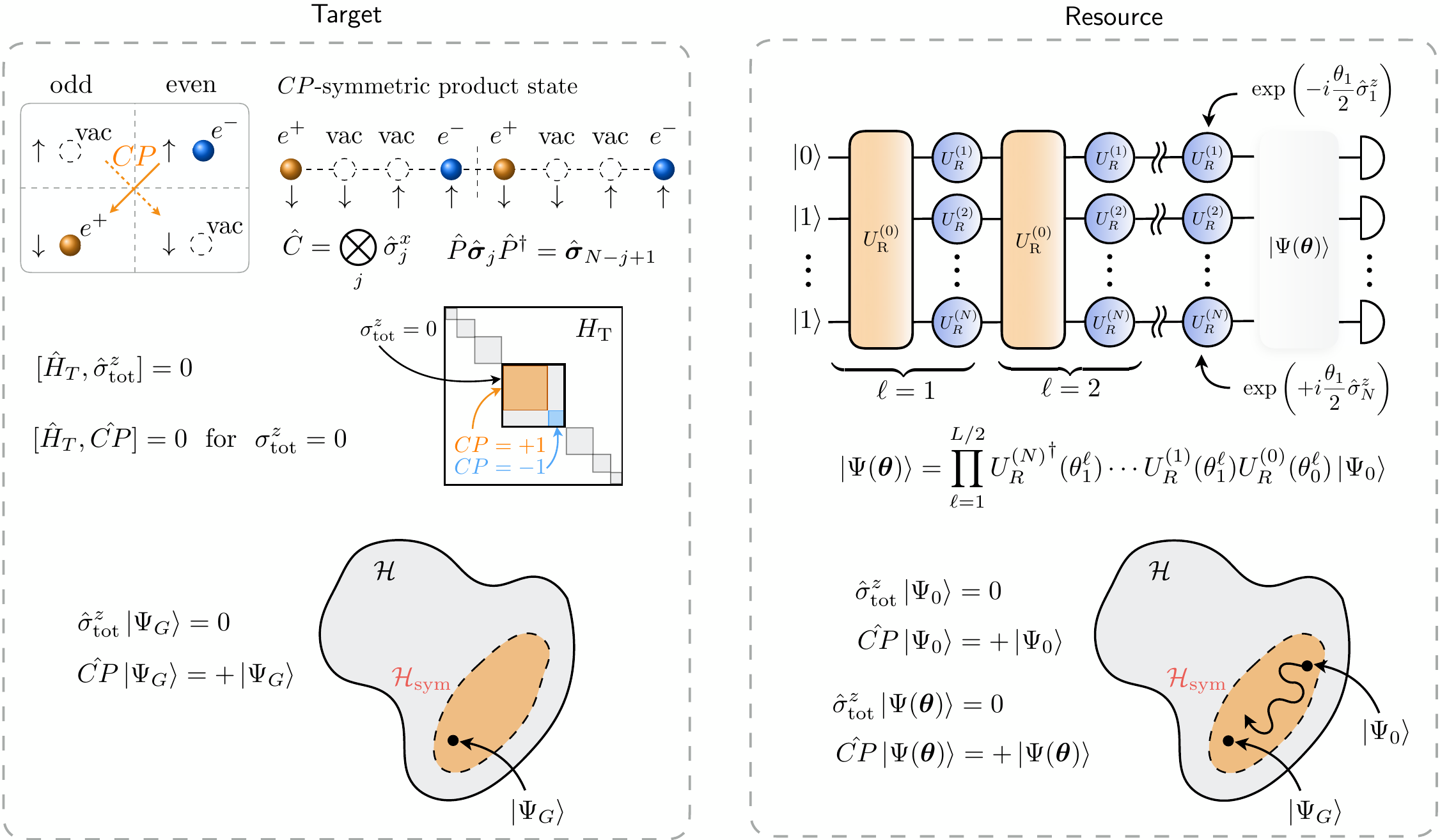}
 \caption{ \label{fig:symmetry}
 {Protection of target model symmetries:}
 (left) Schwinger spin model in the Kogut-Susskind formulation where matter fields are represented by spin degrees of freedom: $\Htarg = 2w \sum_{n=1}^{N-1} \left( \sigma_n^x \sigma_{n+1}^x + \sigma_n^y \sigma_{n+1}^y \right) + \sum_{i=1}^N c_i \sigma_i^z + \sum_{i,j>1} c_{ij} \sigma_i^z \sigma_j^z$. The gauge fields have been eliminated using the Gauss law which results in complicated long-range spin-spin interactions $c_{ij}$ (see Appendix \ref{sec:elimination}). The Schwinger Hamiltonian $\Htarg$ is block-diagonal with respects to different sectors of $\sigma^z_\text{tot}$. Furthermore, the $\sigma^z_\text{tot}=0$ sector decomposes into 2 blocks corresponding to quantum numbers $\CPsym = +1$ and $\CPsym = -1$. We investigate the ground state of $\Htarg$ restricted to the symmetry sector with quantum numbers 0 and $+1$. respectively for the $\sigma^z_\text{tot}$ and the $\CPsym$ symmetries. 
(right) The native resources on an ion trap platform can be exploited to engineer symmetry preserving quantum circuits specifically tailored to the Schwinger model. Taking $B \gg \max\{ |J_{ij}| \}$ (Eq.~\ref{eq:ionIsing}), results in an approximate protection of the $\sigma^z_\text{tot}$ symmetry (see Appendix \ref{sec:elimination}).
Likewise, single-Qubit rotations around the $z$-axis can be forced to be $\CPsym$-symmetric by linking the rotation angles between the left and the right half of the chain according to $\theta^n_{\ell} = - \theta^{N-n+1}_{\ell}$ (see Appendix \ref{App:ProtectingSymmetries}).
Such a circuit will thus protect the target symmetries, restricting the variational search only within the portion of Hilbert space of our interest.
 }
\end{figure*}

The Schwinger model retains, in its spin encoding, the global symmetries of its lattice gauge equivalent \cite{Schwinger_62,Wilson_74,HAMER1982413,Encoding_97}.
Off-diagonal processes consist in the creation or annihilation of an electron-positron pair, thus conserving the (integer) global charge of the system $[\Htarg,\hat{\sigma}^z_{\text{tot}}] = 0$.
This symmetry is a $U(1)$ Lie group $V(\phi) = e^{i \phi \hat{\sigma}^z_{\text{tot}}}$ with Lie generator $\hat{\sigma}^z_{\text{tot}} = \sum_j \hat{\sigma}^z_j$.
The symmetry subspace of our interest is the sector with zero excess charge, which includes the N\'eel state $\ket{\uparrow \downarrow \cdots \uparrow \downarrow}$ corresponding to the bare vacuum, and has quantum number $\hat{\sigma}^z_{\text{tot}} = 0$. Studying this specific symmetry sector allows us to target the quantum phase transition detailed in the Results section.

A more careful consideration must be made when taking into account the $CP$-invariance. Such a $CP$ transformation is the composition of a spatial reflection ($P$) plus a charge conjugation ($C$). In the Kogut-Susskind formalism \cite{KogutSusskind_75}, this operation can be implemented only for a lattice with even length $N \in 2 \mathbb{N}$, as spatial reflection around a bond (not around a site) effectively maps the electron sublattice into the positron sublattice, and vice versa. Since electronic and positronic sites have opposite spin encodings, all the spins must be additionally flipped, i.e. undergo a $\sigma^x$ operation.
The $\CPsym$ transformation thus reads
\begin{equation}
 \CPsym = \prod_{j=1}^{N/2} \left( \hat{\sigma}^x_{j} \hat{\sigma}^{x}_{N+1-j} \hat{W}_{j,N+1-j} \right)
\end{equation}
where $W_{j,j'}$ is the SWAP unitary operator between sites $j$ and $j'$. Overall, $\CPsym$ forms a $Z_2$ discrete group since $(\CPsym)^{\dagger}= (\CPsym)$ and $(\CPsym)^2 = \Id$.
Local spin operators transform under $\CPsym$ as
$(\CPsym) \hat{\sigma}^a_j (\CPsym)^{\dagger} = (\hat{\sigma}^x \hat{\sigma}^a \hat{\sigma}^x)_{N+1-j}$ with $a=x,y,z$. It may be noted that the $\CPsym$ transformation is not a symmetry in the general sense, as it does not leave the full Schwinger Hamiltonian
$\Htarg$ invariant. Indeed, the electron-positron annihilation term is $CP$-invariant itself $(\CPsym) \hat{H}_{T1} (\CPsym)^{\dagger} = \hat{H}_{T1}$,
where $\hat{H}_{T1} = w \sum_j (\hat{\sigma}^+_j \hat{\sigma}^-_{j+1} + h.c.)$,
as well as the bare mass term
$(\CPsym) \hat{H}_{T2} (\CPsym)^{\dagger} = \hat{H}_{T2}$,
where $\hat{H}_{T2} = \frac{m}{2} \sum_j (-1)^j \sigma^z_j$.
The free field term $\hat{H}_{T3} = \bar{g} \sum_{j=1}^{N-1} \hat{L}_j^2$ is the non-invariant component under $\CPsym$, since the dynamical gauge fields $\hat{L}_j = \varepsilon - \frac{1}{2} \sum_{j'=1}^{j} (\hat{\sigma}^z_{j'} + (-1)^{j'})$ transform as
$(\CPsym) \hat{L}_j (\CPsym)^{\dagger} = \hat{L}_j + \frac{1}{2} \hat{\sigma}^z_{\text{tot}}$. Therefore, only within the charge sector $\hat{\sigma}^z_{\text{tot}} \ket{\Psi}= 0$, the gauge fields $\hat{L}_j$, and thus the Hamiltonian component $\hat{H}_{T3}$, are $CP$-invariant, making the $\CPsym$ transformation a subsymmetry for $\Htarg$ in the zero magnetisation sector.
In our simulations, we focus on the even (+) subsector under $\CPsym$, i.e. $\CPsym \ket{\Psi}=  +  \ket{\Psi}$,
which contains the bare vacuum $|\uparrow \downarrow \ldots \uparrow \downarrow \rangle$.

Having characterised the symmetries of the target model, we now implement them on the resources. We specifically tailor a subclass of the resource Hamiltonians $\Hres$ that protects $z$-magnetisation and $\CPsym$ within the zero magnetisation sector (see Fig.~\ref{fig:symmetry}).
The quantum resources we employ consist in an entangling resource Hamiltonian and single qubit rotations. Let us first consider the entangling resource Hamiltonian, which in its native form has long-range Ising-type interactions \cite{PorrasCiracGate,PorrasCiracGateExperiment,MonroePropagation2014,Jurcevic2014}
\begin{equation}
 \Hres^{(0)} = \sum_{i=1}^{N-1}
 \sum_{j=i+1}^{N}
 J_{ij} \hat{\sigma}^x_i \hat{\sigma}^x_j + B \sum_{j} \hat{\sigma}^z_j
\end{equation}
with $J_{ij} \simeq J_0 |i-j|^{-\alpha}$,
and tunable exponent $0 \leq \alpha \leq 3$.
To make it protect $z$-magnetisation, we set $B \gg J_0$. This allows us to adopt degenerate perturbation theory, and approximate
$\Hres^{(0)} \simeq \sum_{i=1}^{N-1}
 \sum_{j=i+1}^{N} J_{ij} (\hat{\sigma}^+_i \hat{\sigma}^-_j + \hat{\sigma}^-_i \hat{\sigma}^+_j) + B \sum_{j} \hat{\sigma}^z_j + O(J_0^2/B)$, which preserves $z$-magnetisation after neglecting the higher order term in $J_0/B$.
 Additionally, $\Hres^{(0)}$ is also $\CPsym$ invariant within the sector $\hat{\sigma}^z_{\text{tot}} = 0$.
 This can be verified by inspecting the $J_{i,j}$ interaction matrix, which is symmetric under reflection  $J_{i,j} = J_{N+1-j,N+1-i}$.
 
 Secondly, we preserve the symmetries for the single qubit resources $\exp(-i  \theta\frac{\Delta_0}{2}\hat{\sigma}^{a}_j )$. In order to preserve the $z$-magnetisation for this term, all the rotations are taken around the $z$ axis, thus $\hat{\sigma}^{a}_j = \hat{\sigma}^{z}_j$. The $\CPsym$ for single-qubit rotations is preserved by matching the site rotation at site $j$ with the rotation at site $(N+1-j)$ site, specifically
 ${ \theta^{(j)} = - \theta^{(N+1-j)}}$.
Tailoring the previous symmetries into the resource gates reduces the effective number of variational control parameters: from $2$ to $1$ for entangling layers, and from $3N$ to $N/2$ for single-qubit layers.


\section{Measurement Scheme}
\label{App:MS}
The closed-loop variational protocol requires the construction of the expectation values $\langle \Htarg \rangle$ and $\langle \Htarg^2 \rangle$ from measurements on the trapped ions.
The experimental platform admits simultaneous projective measurements of all qubits in the logical $z$-basis, via spatially resolved fluorescence \cite{Jurcevic2014}. Single-qubit rotations applied prior to fluorescence detection
enable measurements in different product bases. 
For the 20-ion measurements, we partially correct for decoherence and imperfect initial state preparation by post-selecting on the measurements in the $z$-basis, so retaining only those measurements where the zero-magnetisation is preserved.

{\it Measuring the energy} $-$
%

\begin{figure*}
 \includegraphics[width=0.7\textwidth]{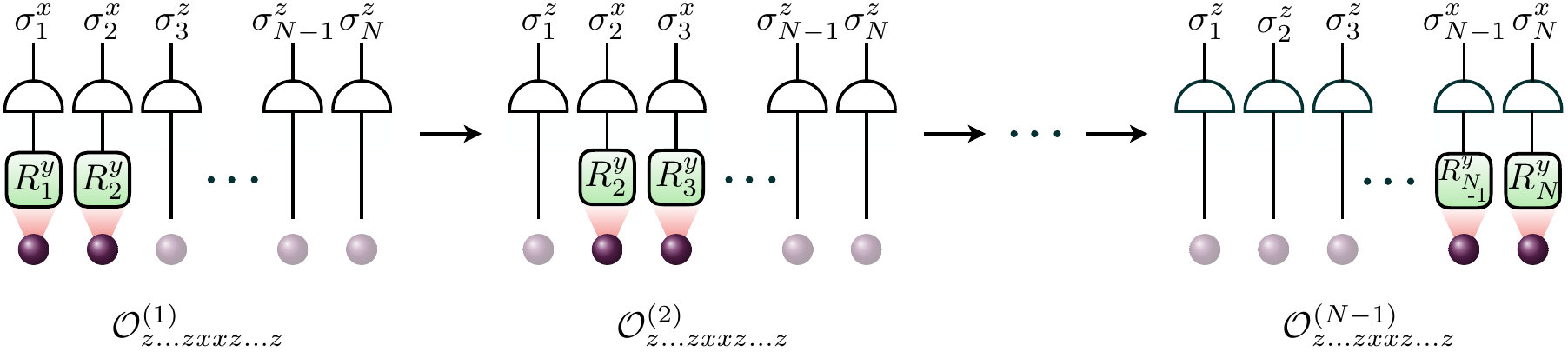}
 \caption{ \label{fig:Measurement}
 {Measurement scheme:}
 Strategy for measuring the squared Schwinger Hamiltonian $\Htarg^2$ from single-qubit operations and measurements. This diagram show the scheme for measuring the remaining components of the anticommutator $\{\Lambda_X, \Lambda_Z \}$ not present in $\Lambda_Y$: They are 4-body correlators of the form $\sigma^x_j \sigma^x_{j+1} \sigma^z_{j'} \sigma^z_{j''}$, with all Pauli operators acting on different sites.
 On the experiment, all such correlators for a specific $j$ are obtained by rotating sites $j$ and $j+1$ under $\exp(i \pi \sigma^{y} /4)$ and then projective measuring in the canonical basis. The procedure has to be repeated for all $j$, thus $N-1$ times.
 }
 \end{figure*}  

The expectation value of the target Hamiltonian $\langle \Htarg \rangle$ is evaluated by decomposing it into three terms
$H= \hat{\Lambda}_X +\hat{\Lambda}_Y +\hat{\Lambda}_Z$, which respectively read
\begin{equation} \label{eq:zedcomponent}
\begin{aligned}
\hat{\Lambda}_X &= \frac{w}{2} \sum_{j=1}^{N-1} \hat{\sigma}_j^x \hat{\sigma}_{j+1}^x, \\
\hat{\Lambda}_Y &= \frac{w}{2} \sum_{j=1}^{N-1}  \hat{\sigma}_j^y  \hat{\sigma}_{j+1}^y, \\
\hat{\Lambda}_Z &= \frac{m}{2}\sum_{j=1}^N (-1)^n \hat{\sigma}_j^z
\\ &\quad+ \bar{g} \sum_{j=1}^{N-1} \left( \varepsilon_0 - \frac{1}{2} \sum_{\ell = 1}^j \left[ \hat{\sigma}_\ell^z + (-1)^\ell \right] \right)^2 \\
 &= \sum_{j=1}^{N} d_j \hat{\sigma}^z_j + \sum_{j =1}^{N-2} \sum_{j' = n+1}^{N-1} c_{j,j'} \hat{\sigma}^z_j \hat{\sigma}^z_{j'}.
\end{aligned}
\end{equation}
These components are measured in the $x$, $y$ and $z$ basis respectively, hence, 3 different measurement bases are needed to evaluate $\Htarg$, regardless of the system size.

{\it Measuring the energy variance} $-$ Estimating the target Hamiltonian variance, $\langle \Htarg^2 \rangle$, requires additional effort. While the square terms $\hat{\Lambda}_X^2$, $\hat{\Lambda}_Y^2$ and $\hat{\Lambda}_Z^2$ can be measured in a single basis each, the anticommutators
$\{ \hat{\Lambda}_Z, \hat{\Lambda}_X \}$ and $\{ \hat{\Lambda}_X, \hat{\Lambda}_Y \}$ need $N-1$ measurement bases each, as we show below in this section.
Therefore, $3N$ different measurement bases are required to evaluate $\langle \Htarg^2 \rangle$.

We now show how to measure the anticommutator observable $\{ \hat{\Lambda}_Z, \hat{\Lambda}_X \}$ in $N-1$ bases. The  $\hat{\Lambda}_Z$ term contains both local and 2-body long range correlations, while $\hat{\Lambda}_X$ contains only short-range correlations. After decomposing the anticommutator term, we observe that some summands disappear by commutation rules, specifically
$\{ \hat{\sigma}^{x}_j \hat{\sigma}^{x}_{j+1}, \hat{\sigma}^z_j \} = \{ \hat{\sigma}^{x}_j \hat{\sigma}^{x}_{j}, \hat{\sigma}^z_{j+1} \} = 0$ and
$\{ \hat{\sigma}^{x}_j \hat{\sigma}^{x}_{j+1}, \hat{\sigma}^z_j \hat{\sigma}^z_{j' \neq j+1} \} = \{ \hat{\sigma}^{x}_j \hat{\sigma}^{x}_{j+1}, \hat{\sigma}^z_{j+1} \hat{\sigma}^z_{j' \neq j} \} =0$.
The surviving terms are of the form
$\hat{\sigma}^x_j \hat{\sigma}^x_{j+1} \hat{\sigma}^z_{j'}$, or $\hat{\sigma}^x_j \hat{\sigma}^x_{j+1} \hat{\sigma}^z_{j'} \hat{\sigma}^z_{j''}$, where each Pauli operator acts on a different site. Additionally, there are parallel 2-site correlators $\hat{\sigma}^a_j \hat{\sigma}^a_{j'}$, which are already known from the energy measurement $\braket{\Htarg}$. For a fixed $j$, we need a single measurement basis to obtain the projective measurement for all 3- and 4-body correlators of the form $\hat{\sigma}^x_j \hat{\sigma}^x_{j+1} \hat{\sigma}^z_{j'}$, or $\hat{\sigma}^x_j \hat{\sigma}^x_{j+1} \hat{\sigma}^z_{j'} \hat{\sigma}^z_{j''}$.
To do so, we need to measure sites $j$ and $j+1$ in the $x$-direction basis, and all the other sites in the $z$-direction basis (see Appendix \ref{App:MS}, Fig.~\ref{fig:Measurement}). Thus, only sites $j$ and $j+1$ are to be rotated by $e^{i \hat{\sigma}^y \pi / 4}$ before the measurement.
To get the whole anticommutator $\{ \hat{\Lambda}_Z, \hat{\Lambda}_X \}$, this procedure has to be repeated for all $j$ from 1 to $N-1$. Thus $N-1$ measurement bases are required.
The same procedure applies to the anticommutator $\{ \hat{\Lambda}_Y, \hat{\Lambda}_Z \}$, where now sites $j$ and $j+1$ should to be rotated by $e^{i \hat{\sigma}^x \pi / 4}$.
Finally, an analogous scheme works for the anticommutator $\{ \hat{\Lambda}_X, \hat{\Lambda}_Y \}$: by rotating sites $j$ and $j+1$ according to $e^{i \hat{\sigma}^y \pi /4}$ and all other sites according to $e^{i \hat{\sigma}^x \pi /4}$ one
obtains the terms of the form $\hat{\sigma}^x_j \hat{\sigma}^x_{j+1} \hat{\sigma}^y_{j'} \hat{\sigma}^y_{j'+1}$ for all $j'$.


It is worth to mention that, in the situation when the magnetisation symmetry is perfectly protected, the measurement scheme becomes even cheaper. Indeed, for a zero-magnetization state, we have $\braket{\hat{\Lambda}_X} = \braket{\hat{\Lambda}_Y}$, hence evaluating $\braket{\Htarg}$ requires only $2$ measurement bases.
Similarly, the anti-commutator terms $\{ \hat{\Lambda}_X, \hat{\Lambda}_Z \}$ and $\{ \hat{\Lambda}_Y, \hat{\Lambda}_Z \}$ are identical, thus the complete procedure to evaluate $\langle \Htarg^2 \rangle$ requires only $2N$ measurement bases.
We remark, however, that the programmable analog quantum simulator we employ, protects the $\hat{\sigma}^z_{\text{tot}}$ symmetry only approximately. More precisely, the term $ B / J_0 \sim 10$ controlling symmetry violations is finite (see Supplementary Information). Moreover, imperfect preparation of the input state $| \Psi_{0} \rangle$ may induce a slight, but measurable, symmetry breaking. We therefore estimate the cost function in $3$ bases for $\langle \Htarg \rangle$ and $3N$ bases for $\langle \Htarg^2 \rangle$.


 \section{Parallelisation via Central Data Repository}
 \label{App:CDR}
A key feature of our VQS scheme, allowing us to reduce the experimental efforts, is the use of a Central Data Repository (CDR). The CDR stores the correlation measurement outcomes produced by the quantum resource and assigns them permanently to their variational control parameters $\paramvec$. This data is processed by the classical computer in order to construct expectation values for a specific target Hamiltonian. 
Since the Hamiltonian parameters only enter at the post-processing stage, we are free to evaluate expectation values of the prepared state for an arbitrary Hamiltonian, as long as it can be measured in the same measurement bases for which the experimental data is available.

We exploit this freedom to speed-up the variational quantum simulation of the phase transition (Sec. \ref{sec:results}, Fig. \ref{fig:qpt}). Here, we first fix the mass $m$ at a certain value and optimise the energy cost function, and store all the measured correlations in the CDR. We then change the value of $m$ to the next point to be optimised, and re-evaluate all the data in the CDR for the new value of $m$. This already provides the optimisation algorithm with a rough outline of the energy landscape even before starting the new optimisation run for the new $m$, and allows us to locate the new global minimum much faster. 



 \section{Energy gap estimation via quantum subspace expansion}
\label{App:QSE}
Here we briefly discuss the technique we employed to provide a quantitative estimate of the (inter-sector) energy gap solely using data from the quantum processor. To do so, we adopt a quantum subspace expansion approximation, put forward in the context of variational quantum eigensolvers \cite{McCleanQSE2017,Colless_2018}.

The method relies, first of all, on having a good approximant of the ground state of the target Hamiltonian. In our case, this is represented by the optimized variational ground state $\ket{\Psi({\paramvec}_{\text{opt}})}$ we obtained from the VQS, with approximate ground energy $E^{(0)} = \bra{\Psi({\paramvec}_{\text{opt}})} \hat{H}_T \ket{\Psi({\paramvec}_{\text{opt}})}$. On top of $\ket{\Psi({\paramvec}_{\text{opt}})}$, we then construct a subspace of symmetry-preserving, low-energy excitations
$\hat{O}_q \ket{\Psi({\paramvec}_{\text{opt}})}$, with $\hat{O}_q$ a selected set of elementary excitation operators.
By performing the corresponding measurements of the variational ground state, it is possible to acquire the effective Hamiltonian matrix 
$H^{\text{eff}}_{q, q'} = \bra{\Psi({\paramvec}_{\text{opt}})} \hat{O}_q \hat{H}_T \hat{O}_{q'} \ket{\Psi({\paramvec}_{\text{opt}})}$
on the subspace, as well as the overlap matrix
$M_{q, q'} = \bra{\Psi({\paramvec}_{\text{opt}})} \hat{O}_q \hat{O}_{q'} \ket{\Psi({\paramvec}_{\text{opt}})}$.
Finding the approximate energy levels is thus cast as the generalised eigenproblem
\begin{equation}
\sum_{q'} H^{\text{eff}}_{q,q'} v^{(k)}_{q'} = \lambda_{(k)} \sum_{q'} M_{q,q'} v^{(k)}_{q'},
\end{equation}
with real eigenvalues $\lambda_{(k)} \leq \lambda_{(k+1)}$. At this stage, $\lambda_{(0)}$ is an improved approximation of the ground state energy and $( \lambda_{(1)} - \lambda_{(0)} )$ is an estimator for the energy gap. The quality of this approach, of course, depends on the quantum subspace chosen for the analysis: Low-energy excitations typically provide substantial corrections in the lower portion of the energy spectrum.

The elementary excitations $\hat{O}_q$ that we consider for constructing the quantum subspace of the Schwinger model include processes of a single, nearest-neighbour, electron-positron pair creation or annihilation. Such processes are the most relevant ones in perturbation theory (for $w \ll g,m$), and
carry an unperturbed energy shift of $(g \pm 2m)$. They also preserve the global charge conservation symmetry. As we additionally want to protect the $\CPsym$ symmetry as well, we tailor these excitations to be explicitly $\CPsym$-symmetric:
\begin{equation}
   \hat{O}_j = \hat{\sigma}^x_j \hat{\sigma}^x_{j+1} + \hat{\sigma}^y_j \hat{\sigma}^y_{j+1}
   + \hat{\sigma}^x_{N-j} \hat{\sigma}^x_{N+1-j} + \hat{\sigma}^y_{N-j} \hat{\sigma}^y_{N+1-j},
\end{equation}
for $j$ from 1 to $N/2$,
while $\hat{O}_0 = \Id$.
According to this prescription, each operator of the form $\hat{O}_j \hat{H}_T \hat{O}_{j'}$ can be decomposed in a sum of Pauli operator strings, and can thus be measured on the quantum simulator via $n$-body correlators. We remark that our prescription for the elementary excitation subspace requires different measurement bases scaling as $\propto N^2$, to reconstruct the full matrices
$H^{\text{eff}}_{j,j'}$ and $M_{j,j'}$. Finally, the generalized eigenproblem itself has simply dimension $N/2 + 1$ and can be solved exactly via standard numerical methods.


\section{Mapping of the Lattice Schwinger Model}
\label{sec:elimination}

\begin{figure}
 \includegraphics[width=\columnwidth]{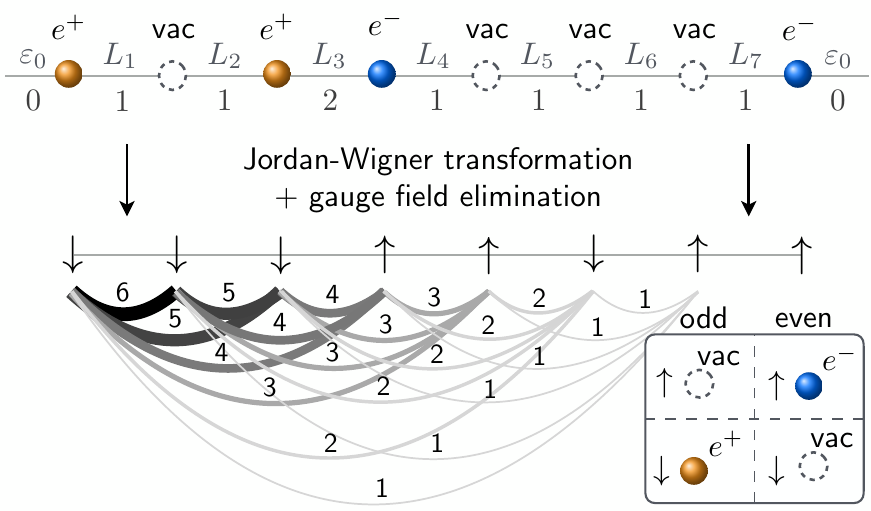}
 \caption{  \label{fig:SchwingerEncoding}
 %
 %
 { Qubit chain encoding of the lattice Schwinger model:}
Illustration of a specific product state for 8 lattice sites in the Schwinger model and its encoding in the corresponding spin configuration. The curves between sites represent the long-range spin-spin interaction pattern, where the thickness of each curve encodes the correspondent coupling strength, corresponding to the couplings $c_{i,j}$ in Eq.~\eqref{eq:zedcomponent}.
The lower box shows again the encoding of particles into spins.
}
\end{figure}

The Schwinger model describes a one-dimensional gauge field theory, where matter is represented by a flavourless fermion field $\Psi(x)$, and
interacts via an Abelian $U(1)$-symmetric gauge field.
In this section,  we briefly summarise the mapping from the quantum electrodynamics model into a spin lattice model with long-range interactions (c.f.~Appendix \ref{sec:elimination}, Fig. \ref{fig:SchwingerEncoding}), which is the starting point of VQS in the present paper 
\cite{HAMER1982413,Encoding_97,Muschik2017,Banuls2013a,Martinez_2016}.
In the temporal gauge, where the temporal component of the vector potential is set to zero, $A_0(x) = 0$, the Hamiltonian density of the continuum model reads
\begin{multline}
\hat{H} =\int dx \left[ {\Psi}^{\dagger}(x)\gamma^0 \gamma^1\left(-i\partial_1+ g\hat{A}_1(x)\right){\Psi}(x)
\right. \\ \left.
+m {\Psi}^{\dagger}(x)\gamma^0 \Psi(x)+\frac{1}{2}\hat{E}^2(x)\right],
\end{multline}
where the longitudinal vector potential $\hat{A}_1$ and the electric field $-\hat{E}(x) = \partial_0 \hat{A}_1(x)$ satisfy the commutation relation
$[\hat{A}_1(x),\hat{E}(x')]=-i\delta(x-x')$. The Gauss' law $\partial_1 E = g \Psi^{\dagger} \Psi$ is the local $U(1)$ gauge symmetry of the model.
In 1D, the Dirac matrices read $\gamma^0=\hat{\sigma}^z$ and $\gamma^1=i\hat{\sigma}^y$.
The Kogut-Susskind formulation allows the field theory to be recast on a lattice, with lattice constant $a$.
The gauge fields $\theta_{j,j+1} = -ag A(x_{j}+\frac{a}{2})$
and $L_{j_j+1} = \frac{1}{g} E(x_{j}+\frac{a}{2})$
now live on the lattice bonds $\{j,j+1\}$, and still canonically commute $[\hat{\theta}_{j,j+1},\hat{L}_{j',j'+1}]=i\delta_{j,j'}$.
The fermion fields are now spinless, since the two spinor components of the theory decouple, and live on the lattice sites:
$\hat{\Phi}_{2j}=\sqrt{a}\hat{\Psi}_{e^{-}}(x_{2j})$ for even sites and 
$\hat{\Phi}_{2j-1}=\sqrt{a}\hat{\Psi}^{\dag}_{e^{+}}(x_{2j-1})$ for odd sites.
With these prescriptions, the resulting lattice model thus reads

\begin{multline}
\label{Eq_LatticeHamiltonian}
\hat{H} =  \frac{1}{2a} \sum_{j=1}^{N-1}\left[\hat{\Phi}^{\dag}_j \hat{U}_{j,j+1} \hat{\Phi}_{j+1} + h.c.\right]
\\
+m \sum_{n=1}^{N}(-1)^j \hat{\Phi}^{\dag}_j\hat{\Phi}_j + \frac{g^2 a}{2} \sum_{j=1}^{N-1} \hat{L}_j^2,
\end{multline}
where $\hat{U}_{j,j+1} = e^{i \theta_{j,j+1}}$ obeys the commutation relations
$[\hat{U}_{j,j+1}, \hat{U}^{\dagger}_{j',j'+1}] = 0$ and $[\hat{L}_{j,j+1}, \hat{U}_{j',j'+1}] = i \delta_{j,j'} \hat{U}_{j,j+1}$.
Here, open boundary conditions have been assumed, and $N$ is the system length.
In this formalism, the Gauss' law reads
$\hat{L}_{j,j+1} - \hat{L}_{j-1,j} = \hat{\Phi}^{\dagger}_j \hat{\Phi}_j - \frac{1-(-1)^j}{2}$.

According to this Gauss' law, the fermion configuration and the gauge field at the (left) boundary $\hat{L}_{0,1} = \varepsilon_0$ determine completely
the gauge fields' configuration. It is thus possible to {\it eliminate} the gauge fields, so that they are no longer independent dynamical degrees of freedom, but collective operators on the matter degrees of freedom, specifically  as $\hat{L}_{j,j+1} = \varepsilon_0 + \sum_{\ell = 1}^{j} \left[ \hat{\Phi}^{\dagger}_j \hat{\Phi}_j - \frac{1}{2}(1-(-1)^{\ell}) \right]$. Similarly, the $\hat{U}_{j,j+1}$ are removed from the Hamiltonian by a gauge transformation of the fermi fields, namely
$\hat{\Phi}_j \to (\prod_{\ell=1}^{j-1} \hat{U}^{\dagger}_{\ell,\ell+1} ) \hat{\Phi}_j$. Such transformations lead to a resulting model where the fermionic hopping is nearest-neighbour, while the density-density interactions are long range:
\begin{multline}
\label{Eq_LatticeHamiltonian2}
\hat{H} =  w \sum_{j=1}^{N-1}\left[\hat{\Phi}^{\dag}_j \hat{\Phi}_{j+1} + \text{h.c.}\right]
+m \sum_{n=1}^{N}(-1)^j \hat{\Phi}^{\dag}_j\hat{\Phi}_j +
\\
+\bar{g} \sum_{j=1}^{N-1} \left( \varepsilon_0 + \sum_{\ell = 1}^{j}  \left[ \hat{\Phi}^{\dagger}_j\hat{ \Phi}_j - \frac{1-(-1)^{\ell}}{2} \right] \right)^2,
\end{multline}
where $w = \frac{1}{2a}$ and $\bar{g} = \frac{g^2 a}{2}$. As a final manipulation, we apply a Jordan Wigner transformation, mapping
fermions into spins according to $\Phi^{\dagger}_j = \prod_{\ell < j} (-\hat{\sigma}^z_{\ell}) \hat{\sigma}^{+}_j$. This mapping recovers a Hamiltonian 
of the form
\begin{multline} \label{eq:schwinger_ugly}
\Htarg = w \sum_{j=1}^{N-1} \left[ \hat{\sigma}_j^+ \hat{\sigma}_{j+1}^- + \text{h.c.}  \right] 
+ \frac{m}{2}\sum_{j=1}^N (-1)^j \hat{\sigma}_j^z 
\\
+ \bar{g} \sum_{j=1}^{N-1} \left( \varepsilon_0 - \frac{1}{2} \sum_{\ell = 1}^j ( \hat{\sigma}_\ell^z + (-1)^\ell ) \right)^2,
\end{multline}
corresponding to the spin model of the main text (using $w=1$).

\section{Variational quantum simulator workflow}
\label{App:VQSworkflow}
In the present work, we adapt a variational quantum eigensolver
\cite{MollGambetta2018,Farhi2014,Kandala2017,Otterbach_2017,OMalley2016,Farhi2016}
analogous to those employed in the previous years {\it e.g.}~for quantum chemistry
\cite{Peruzzo2014,Rohringer2008,VQATroyer2015,VQAMcClean2016,VQSPontryagin2017,Hempel_2018,AspuruGuzikReview2018}.
Here we employ it to study intrinsically-scalable lattice models as a target models.
The goal of this algorithm is to prepare and optimise the ground state
of a target hamiltonian $\Htarg$, which we decompose into $n$-body interaction terms
\begin{equation} \label{eq:ht}
\Htarg =  h_k^{\alpha} \, \hat{\sigma}_{k}^{\alpha} +  h_{k \ell}^{\alpha \beta} \, \hat{\sigma}_{k}^{\alpha} \hat{\sigma}_{\ell}^{\beta} + h_{k \ell n}^{\alpha \beta \gamma} \,  \hat{\sigma}_{k}^{\alpha} \hat{\sigma}_{\ell}^{\beta} \hat{\sigma}_n^{\gamma} + \cdots\,,
\end{equation}
or simply $H =  \sum_{q} h_q \hat{\Gamma}_q$, where the $\hat{\Gamma}_q$ are the Pauli operator strings appearing in the decomposition \eqref{eq:ht}.
Entangled, variational quantum states $\ket{\Psi(\paramvec)}$ are prepared on a quantum simulator, starting from a simple product state $|\Psi_0\rangle$ to which a sequence of unitaries $\exp(- i \theta_m \Hres^{(i_m)})$ is applied. $\{ \Hres^{(j)} \}$ is the set of controllable resource Hamiltonians of our programmable analog quantum simulator.
Optimizing the state preparation is thus cast into the minimisation of the cost function
$\mathcal{F}(\paramvec) = \bra{\Psi(\paramvec)} \Htarg \ket{\Psi(\paramvec)}$. 
A detailed flowchart of VQS is described as follows:
\begin{enumerate}
 \item A set of classical control variables $\paramvec = \{ \theta_k \}$ are passed by a classical computer to the controllable quantum device.
 \item The state $\ket{\Psi_0}$ is initialised on the quantum device. The controllable dynamics, determined via
 $\paramvec$, is performed on the programmable analog quantum simulator and the variational quantum state $|\Psi({\paramvec})\rangle = \exp(- i \theta_k \Hres^{(i_k)}) \cdots \exp(- i \theta_1 \Hres^{(i_1)}) |\Psi_0\rangle$ is prepared for a resource circuit of fixed depth $k$.
 \item The prepared state $|\Psi(\paramvec)\rangle$ is measured in a specific projective measurement basis.
 This provides a single sample measurement $M_s[\hat{\Gamma}_q] = M_x[ \hat{\sigma}_{k}^{\alpha} \hat{\sigma}_{\ell}^{\beta} \hat{\sigma}_n^{\gamma} \cdots]$ for all the components ($q$) compatible with the selected measurement basis. The measurement is stored on the
 Central Data Repository (see Appendix \ref{App:CDR}).
 \item Steps 2.~and 3.~are repeated for different measurement bases, until one sample
 $M_s[\hat{\Gamma}_q]$ has been acquired for every component $q$ of the target Hamiltonian.
 \item Steps 2.,~3.,~and 4.~are repeated multiple ($\bar{s}$) times to reconstruct sample averages
 $\langle \hat{\Gamma}_q \rangle = \bar{s}^{-1} \sum_{s}^{\bar{s}} M_s[ \hat{\Gamma}_q]$ and their related shot-noise variances
$\Delta \hat{\Lambda}^2_{q} = \bar{s}^{-1}  \sum_{s}^{\bar{s}} (M_s[ \hat{\Gamma}_q] - \langle \hat{\Gamma}_q \rangle)^2$.
At this stage, $\bar{s}$ can be specifically increased for those components $q$ that are contributing with the largest variance $\Delta \hat{\Lambda}^2_{q}$.
\item The cost function $\mathcal{F} = \sum_q h_q \langle \hat{\Gamma}_q \rangle$ and its statistical error bar, defined as
 $\Delta \mathcal{F}= ( \sum_q h_q \Delta \hat{\Lambda}^2_q )^{1/2}$ are evaluated.
 \item The classical computer employs the cost function $\mathcal{F}(\paramvec)$ and its statistical error bar $\Delta \mathcal{F}(\paramvec)$ within a global
 optimisation algorithm (see Appendix \ref{App:OA}), capable of minimizing noisy functionals and equipped with the option of re-exploring previously-attempted landscape points $\paramvec$ in order to improve their statistical error bar.
 The algorithm thus provides a new set of control variables $\paramvec$
 and restarts from step 1. The cycle then continues until the resource budget, in terms of single sample measurements, is reached (or a convergence threshold is reached first).
 \end{enumerate}
 

\section{On Scalability}

\begin{figure*}
 \includegraphics[width=0.7\textwidth]{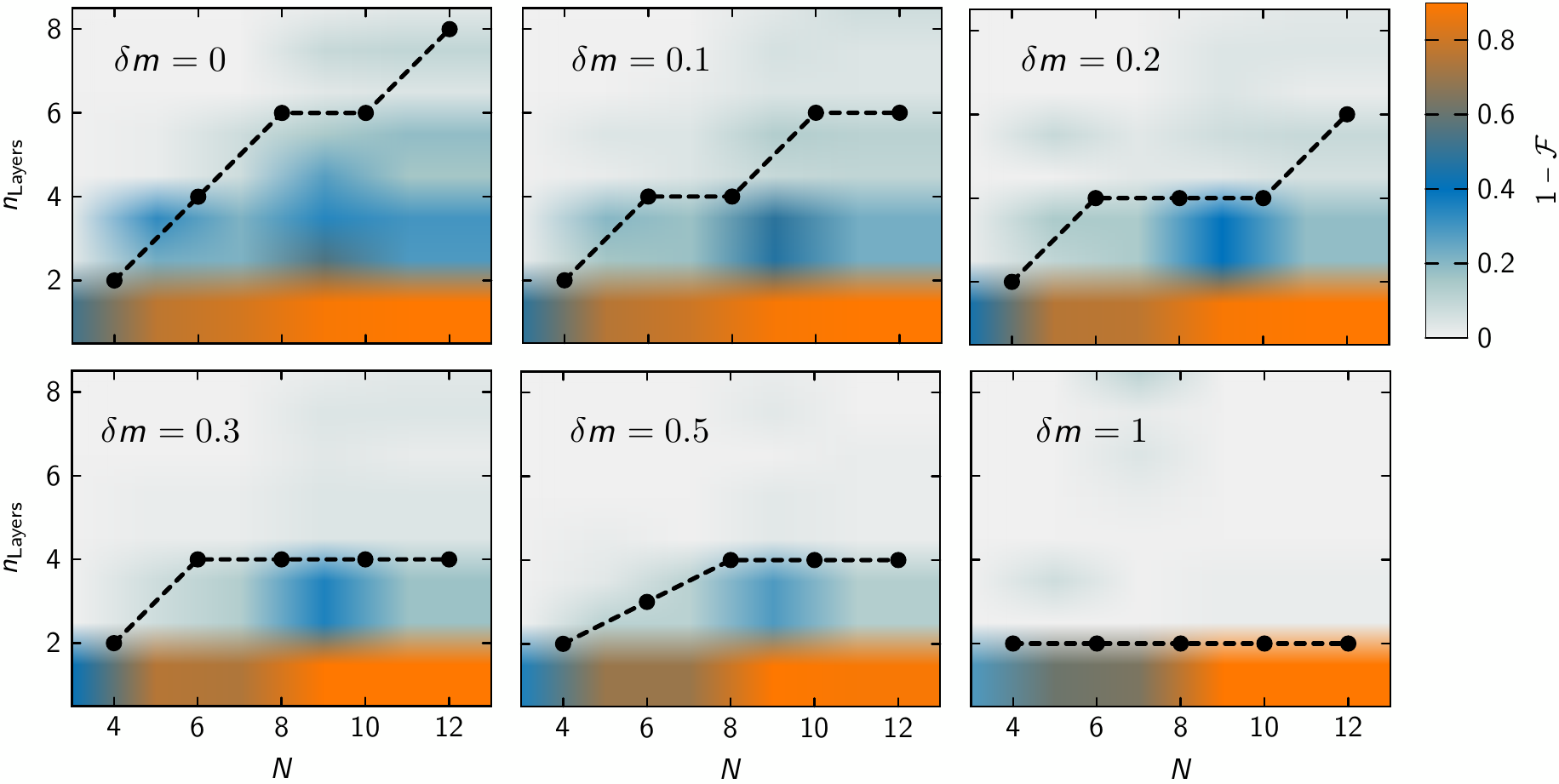}
 \caption{ \label{fig:Scalability}
 { Theoretical scalability}
 Numerical simulation of the VQS for Schwinger ground states with trapped-ion resources. The infidelity of the optimised ground state with respect to the exact ground state is plotted as a function of the number of qubits $N$ and the circuit depth, or number of circuit layers $n_{\text{layers}}$. Different panels refer to the simulation of the Schwinger model for different values of the bare mass $m$, reported via the parametric distance $\delta m = (m - m_c)$ from the critical point $m_c$ (for $w\!=\!\bar{g}\!=\!1$).
 The white curve marks the minimal circuit depth required to achieve an optimised infidelity of 5\% or less, as a function of the system length $N$.
 }
 \end{figure*}
 
 We now address questions related to the complexity and scalability \cite{OMalley2016,OberthalerQSim2018,Santagati2018,DirectConvergence2004,LloydMontangero2014}
 of the VQS algorithm we employ, applied to the specific example of Schwinger model ground states simulated with trapped-ions. We performed and collected numerical simulations of the VQS, investigating the variational complexity required to prepare the ground state within a specified precision. In our VQS framework, the circuit depth is a good quantifier of the variational complexity, as the number of real variational parameters $\paramvec$ scales linearly with the number of circuit layers $n_{\text{layers}}$ at fixed size $N$. For numerical simulations, we can quantify the precision of the variational ansatz by calculating the infidelity of the variationally optimised state
 with the exact ground state of the model, which can be obtained via exact diagonalisation for small system sizes (we considered system sizes $N$ up to 12 sites). We observed clear patterns in the minimal infidelity achievable by the VQS, as a function of the system size $N$, the circuit depth $n_{\text{layers}}$, and the parametric distance of the critical point $\delta m = (m - m_c)$ of the bare mass coupling in the Schwinger model $\Htarg(m)$ we are targeting (see Fig.\ref{fig:Scalability}). For any specific desired precision in the state preparation, {\it e.g.} $5\%$ infidelity, we can identify a minimal circuit depth $n_{\text{req}}$ required to prepare the VQS state at such precision.
 From our simulations, we see that the required layers $n_{\text{req}}$ are a smooth function of $N$ and $\delta m$. More precisely, $n_{\text{req}}$ tends to saturate at finite values, while increasing $N$, away from the critical point. Conversely, close to the critical point $\delta m \sim 0$, $n_{\text{req}}$ is well approximated by a linear function in $N$. Such behaviour is compatible to the analysis we performed directly on the experiment, carried out for 8 sites and variable circuit depth (Fig.~\ref{fig:8IonsAlgoError}).
 
 This emergent scaling behaviour suggests that the variational complexity (at finite algorithmic error) of the Schwinger ground state is polynomial, and not exponential, in the system size $N$, ultimately making our VQS procedure scalable (assuming a polynomially-scaling search algorithm is used). Such a result is compatible with notions from Tensor Network theory \cite{Rico2014,BuyensDMRGQED,Banuls2013a,Banuls2013b}. Even though the Schwinger model has long-range interactions, its ground state satisfies one-dimensional area laws of entanglement (or violates them logarithmically in $N$ at the critical point). This observation follows from the fact that the Schwinger model is short-range interacting prior to the elimination of the dynamical gauge field (see Sec.~\ref{sec:elimination}), and the elimination itself occurs without an increase in entanglement content. In turn, this tells us that the Schwinger ground state, at finite algorithmic error, carries a polynomial variational complexity, i.e. it can be encoded into a Matrix Product State with bond dimension growing polynomially in $N$.
 Recent studies analysing scalability of hybrid feedback-loop simulations also corroborate these predictions \cite{LloydMontangero2014,Doria2011,VanFrank2016}.
 
 \section{On Controllability and Symmetries}
 
 In this section we review some elements from controllability theory, and analytically discuss how to relate resource symmetries with the targetability of a many-body quantum state via VQS in the limit of infinite resources \cite{controllability83,LloydMontangero2014}. Specifically, we will assume that the quantum simulator experimental platform we employ, the programmable analog quantum simulator, is non-universal, but we can dedicate an arbitrary amount of resources, in terms of circuit depth, to it. We will study which quantum states can be achieved and which target Hamiltonians can be effectively simulated.
 
 First of all, let us introduce the Lie algebra $\mathcal{L}$ of resource Hamiltonians. This is the closure under (real) linear combinations and commutation (times $i$) of the bare resource Hamiltonians, and describes the most general dynamics available on the platform. To show this property, we will now prove that if $\hat{A}$ and $\hat{B}$ are resource Hamiltonians, so are $\lambda \hat{A} + \hat{B}$ and $i[\hat{A},\hat{B}]$. In the following, we will assume time-reversal $e^{i t \Hres} \to e^{- it \Hres}$ is available for simplicity of the proofs.
 \\
 {\it Combination} $-$ Let us consider the following circuit with 2k+2 circuit layers, and resource Hamiltonian $\hat{A}$ and $\hat{B}$:
 \begin{equation}
     \left( e^{\frac{i \lambda t}{2 k} \hat{A}} e^{\frac{i t}{2 k} \hat{B}} e^{\frac{i \lambda t}{2 k} \hat{A}}\right)^k = e^{i t (\lambda \hat{A} + \hat{B})} + O( (t/k)^3),
 \end{equation}
 where the equality follows from the second-order Suzuki-Trotter formula. In the unlimited resource limit $k \to \infty$, the equality becomes exact, thus $\lambda \hat{A} + \hat{B}$ is a resource Hamiltonian. \\
 {\it Commutation} $-$ As before, we consider resource Hamiltonians $A$ and $B$, and the circuit with 4k layers:
 \begin{equation}
     \left( e^{\frac{i t}{k} \hat{A}} e^{\frac{i t}{k} \hat{B}} e^{- \frac{i t}{k} \hat{A}} e^{-\frac{i t}{k} \hat{B}} \right)^k = 
     e^{i t^2 ( i [\hat{A},\hat{B}] )} + O( (t/k)^3),
 \end{equation}
  where the equality follows from the Baker-Campbell-Hausdorff formula. In the limit $k \to \infty$ of infinite resources, we can exactly generate the resource Hamiltonian $i[\hat{A},\hat{B}]$.
  
Next, we argue that if two states are dynamically disconnected, then it is the result of a symmetry present in the resources. To this end, assume we have $|\phi_1\rangle$ and $|\phi_2\rangle$ such that $\langle \phi_1 | e^{i \Hres t} | \phi_2 \rangle = 0$ for all $H_R$ in the resource Lie algebra $\mathcal{L}$. It follows that the group of resource unitaries $e^{i \Hres t}$ forms a reducible representation, and that $|\phi_1\rangle$ and $|\phi_2\rangle$ belong to two different irreducible subspaces, $\mathcal{S}_1$ and $\mathcal{S}_2$ respectively, orthogonal and with nonzero dimension. Now we consider the unitary operator $\hat{Q} = \Id - 2 \hat{P}_1$, where $\hat{P}_1$ is the projector onto subspace $\mathcal{S}_1$. $Q$ is a nontrivial operator and forms a $\mathbb{Z}_2$ group via $\hat{Q}^2 = \Id$.
Moreover, for any state $|\Psi\rangle$ in $\mathcal{S}_1$ we have $e^{i \Hres t} \hat{Q} |\Psi\rangle = - e^{i \Hres t} |\Psi\rangle = \hat{Q} e^{i \Hres t} |\Psi\rangle$ since $e^{i \Hres t} |\Psi\rangle$ still belongs to $\mathcal{S}_1$. Similarly, for states $|\Psi_{\perp}\rangle$ in the orthogonal of $\mathcal{S}_1$, we have $e^{i \Hres t} \hat{Q} |\Psi\rangle = + e^{i \Hres t} |\Psi\rangle = \hat{Q} e^{i \Hres t} |\Psi\rangle$. We conclude that $[\Hres,\hat{Q}] = 0$ for any $\Hres \in \mathcal{L}$, which means that $\hat{Q}$ is a symmetry for the resources.

The previous argument encourages us to characterise the resources in terms of their symmetries $\hat{V} \in \mathcal{G}$ and their group $\mathcal{G}$, containing all unitaries $\hat{V}$ commuting with the resource Lie algebra $[\Hres, \hat{V}] = 0$  $\forall\, \hat{V} \in \mathcal{L}$.
The state preparation during VQS will, of course, preserve the symmetry content (in terms of quantum numbers and conserved Noether currents) \cite{NoetherInvariant}, thus it is mandatory that the quantum many-body state we are targeting exhibits the same symmetry content.

Since our target states are eigenstates of a target Hamiltonian $\Htarg$, we know that they will indeed possess good quantum numbers if $\Htarg$ is also invariant under $\mathbb{G}$ - that is - if $\Htarg$ possesses the same symmetries of the resources. Thus, the target Hamiltonian must contain all the resource symmetries to guarantee controllability.

The upside of this consideration is that, for a target Hamiltonian with additional symmetries than the resources, it is possible, depending on the experimental architecture, to tailor these additional symmetries onto the resources. As in the case of VQS of the Schwinger model with trapped-ions, this generally provides a reduction of the variational parameters per circuit layer.

\section{Experimental details}
\label{App:ExperimentalDetails}
In this section, we provide further information about the experimental system that is used as a quantum resource for the variational quantum simulation. The system consists of a string of $^{40}$Ca$^+$ ions confined in a linear Paul trap. A qubit is encoded in two long-lived electronic levels, $\ket{\downarrow} = \ket{S_{1/2},m_j=1/2}$ and $\ket{\uparrow} = \ket{D_{5/2},m_j=5/2}$, of each calcium ion, coupled through an optical quadrupole transition at 729~nm. The experimental sequence for the quantum simulation is divided into three main steps: the first step consists in the preparation of a N\'eel state as an initial state for the algorithm. In the second step, the time evolution of the initial state is carried out for a given set of control variables $\paramvec = \{ \theta_k \}$ that is supplied by a classical computer to the quantum system. The third step is the detection of the quantum state in the required bases after which the measured values are passed to the classical computer. 

{\sl Initialisation:} The initial state is prepared by first Doppler cooling the ion chain for 3~ms, followed by optical pumping for 200~$\mu$s and subsequent sideband cooling of the ion chain. Finally, a set of laser pulses induces light shifts on individual ions in order to prepare the N\'eel state. The duration of the sideband cooling step (6~ms (11~ms) for eight (twenty) ions) is chosen such that all the radial motional modes are cooled close to the ground state. For each optimisation run, the initial state is prepared with high fidelity - for an 8-ion chain, the N\'eel state fidelity is measured to be 98(2)\% and for the 20-ion N\'eel state 91(3)\%. A drop in fidelity is to be expected when increasing the length of the ion chain, due to the accumulation of errors from each ion in the system.

{\sl Trial state preparation:} The second step of the experimental sequence consists of single-qubit rotations and entangling gate operations. Single-ion addressing is implemented by an off-resonant 729~nm laser beam (waist of $\approx 2~\mu$m) steered over the ion chain using an acousto-optic deflector. This generates local light shifts, enabling single-qubit rotations about the $z$-axis \cite{Jurcevic2014}. A bichromatic laser beam generates an XY Hamiltonian with a coupling range {$\alpha = 1.34$} for an 8-ion chain, and {$\alpha = 0.98$} for a 20-ion chain. This coupling range is achieved by choosing the splitting of the bichromatic laser frequencies to be $\pm( \omega_{com} + \delta$), where $\omega_{com} =2 \pi \cdot 2.71$~MHz is the highest radial mode frequency and {$\delta = 2 \pi \cdot 0.04$~MHz}. Additionally, the mean frequency of the bichromatic laser is additionally shifted by $\delta_B = 2 \pi \cdot 3$~kHz from the qubit transition frequency, generating an effective magnetic field in the system.  

{\sl Measurement:} In order to make projective measurements in the required bases, i.e. 3 (24) bases for the energy (variance) estimation, the corresponding single-qubit eigenstates are aligned with the $z$-direction by a sequence of coherent operations composed of single-qubit gates sandwiched between resonant $\pi/2$-pulses that globally rotate all qubits by the same amount \cite{LanyonTomography2017}. Subsequently, the quantum state of each individual ion is read-out using a spatially resolved fluorescence detection scheme. The fluorescence detection involves the interrogation of the ions using a 397~nm laser beam for 1~ms, with the scattered photons detected using an EMCCD camera. 
For the 20-ion measurements, we partially correct for decoherence and imperfect initial state preparation by post-selecting on the measurements in the $z$-basis. Therefore, we retain only those $70\%$ of the measurements, which lie within the zero-magnetisation subspace.

The entire experimental sequence is repeated for between 30 and 200 iterations, and the statistical average taken. The data is subsequently passed to the classical computer, with the classical-quantum feedback loop running in a fully automated manner. The system is initially calibrated for a given number of ions and prepared for the quantum simulation. The classical computer then generates a script file containing all the aforementioned experimental steps. The experimental control system processes the script file and generates the required laser sequence, which is then applied to the trapped ions. Subsequently, the projective measurements are performed and the data is stored in the CDR (see Appendix \ref{App:CDR}). The experimental control system also calibrates itself against any slow drifts of the experimental parameters such as laser frequency, the positioning of the ion chain, and drift of the addressing laser beam. The calibration sequence is repeated every 10 minutes to correct for the drifts, and appropriate adjustments are consequently carried out.

\section{Figures: Algorithmic error analysis and 16 ion optimisation}
\label{App:Fig}
\begin{figure}
 \includegraphics[width=\columnwidth]{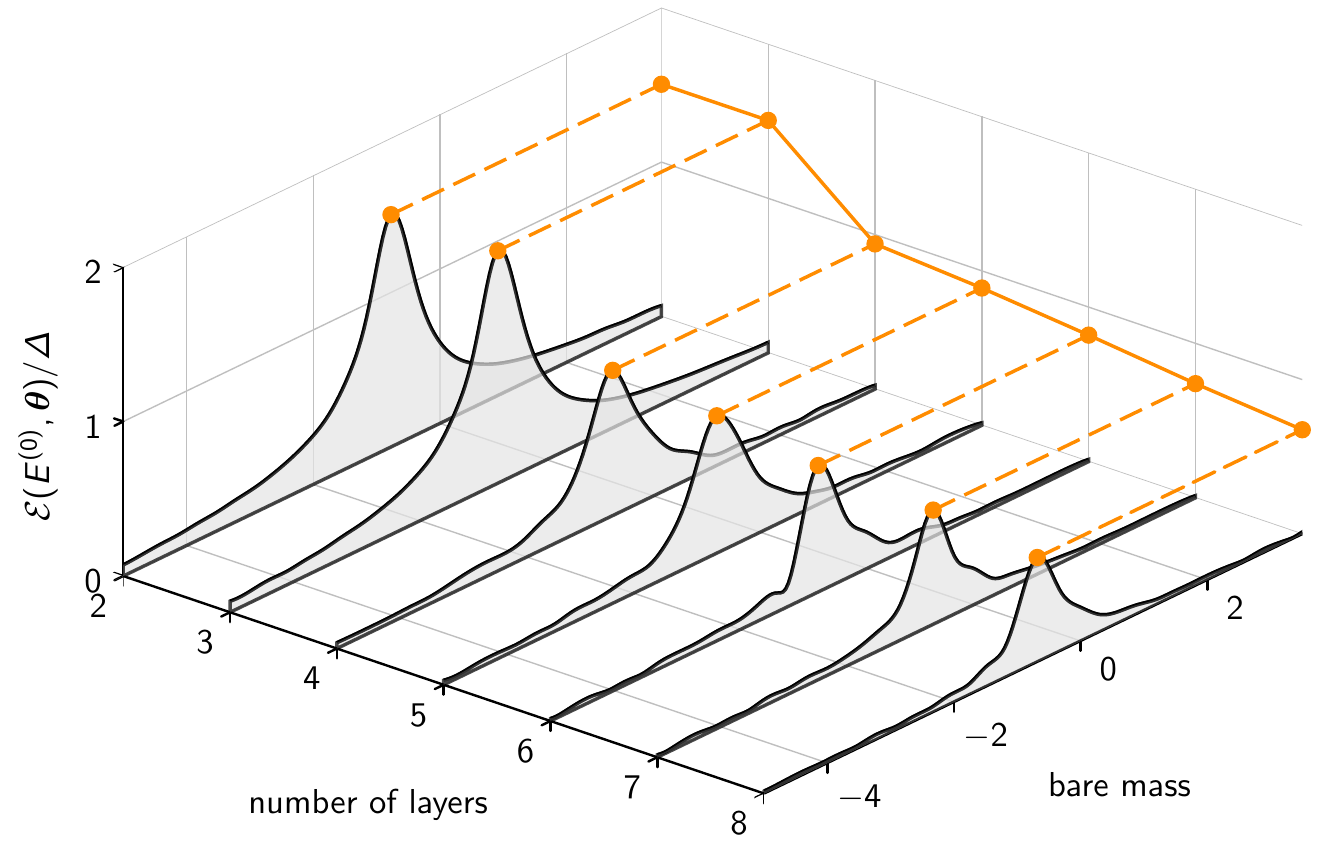}
 \caption{ \label{fig:8IonsAlgoError}
 { Analysis of the algorithmic errorbar:} Theoretically expected standard deviations of the target Hamiltonians, obtained from a numerically simulated experiment optimising the ground state for 8 ions, are plotted as a function of the bare mass $m$ around the critical point, and of the circuit depth
 (with $w = \bar{g} = 1$). As expected, algorithmic error bars decrease for increasing circuit depth. This is especially visible in proximity of the critical point (peaks in the black curves, also reported as orange dots) where the target states are more entangled, requiring deeper circuits to achieve a required precision.
 }
\end{figure}

\begin{figure*}
 \includegraphics[width=0.95\textwidth]{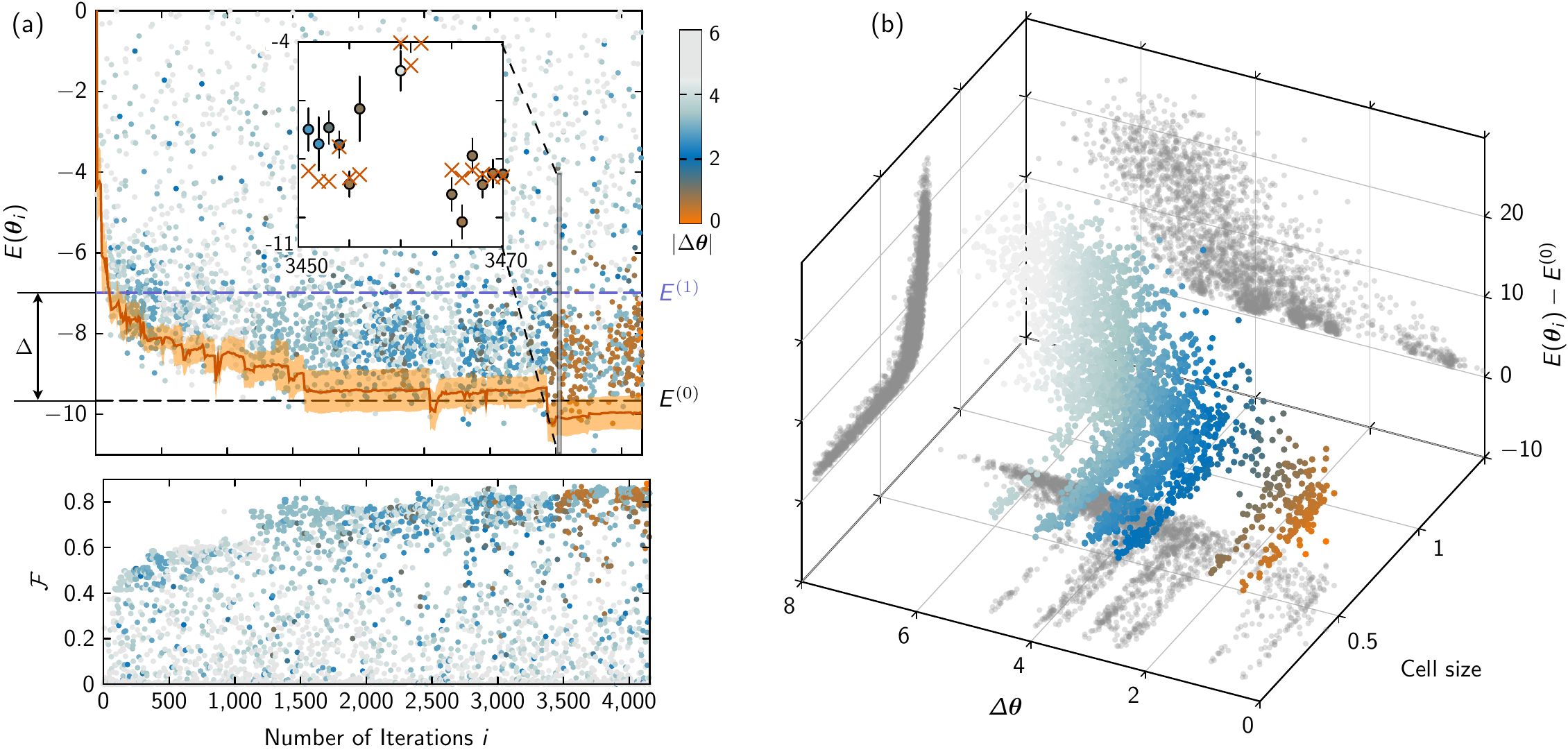}
 \caption{ \label{fig:16ions}
 { (a) Energy minimisation for 16 ions:} (upper panel) Experimentally measured
energies $E^{(0)}(\paramvec_i)\equiv\bra{\Psi(\paramvec_i)}\Htarg\ket{\Psi(\paramvec_i)}$ (dots) in the course of a single optimisation run for 16 ions, plotted
vs.~iteration number $i$ of the DIRECT optimisation algorithm (see text), for  $m\!=\!0.6,w\!=\!\bar{g}\!=\!1$. Energy values $E^{(0)}({\paramvec_i})$  are colour-coded to indicate the Euclidian distance of $\paramvec_i$ to the final optimised parameter vector $\paramvec_\text{opt}$, as selected by theoretical fidelity. The solid red line indicates the algorithm's current estimate of the groundstate energy and its $2\sigma$ uncertainty (shaded area), from modelling the thus far observed energies as jointly gaussian distributed random variables (see Appendix \ref{App:OA}). Inset: Close-up of a late stage of the optimisation, where statistical errorbars (see Appendix \ref{App:VQSworkflow}) are displayed, and theoretically simulated values are plotted as crosses. The lower panel displays theoretically calculated fidelities $\mathcal{F}$ corresponding to the experimentally applied parameters $\paramvec_i$.
 { (b) Visualisation of the energy landscape.}
Sampled energies plotted as a function of their distance in the 15-dimensional parameter space $|\Delta \Theta|$ relative to the optimal point $\paramvec_\text{opt}$, and the cell size that each sampling point represents in the DIRECT algorithm (see Appendix \ref{App:OA}). The algorithm encountered several local minima, appearing as distinct branches with low energies at specific parameter distances and extending towards the direction of smaller representative cell sizes, which is indicative of an increasingly dense sampling around each of the local minima.
 }
\end{figure*}

Figures \ref{fig:8IonsAlgoError} shows an analysis of the algorithmic error as a function of circuit depth and bare mass. Figure \ref{fig:16ions} shows an optimisation run for 16 ions with a large budget. The algorithm encountered several local minima in the energy landscape.

\clearpage

\bibliographystyle{apsrev4-1}

\end{document}